\documentclass[english,superscriptaddress]{revtex4-2}
\usepackage[T1]{fontenc}
\usepackage[latin9]{inputenc}
\usepackage{amsmath}
\usepackage{graphicx}
\usepackage{babel,comment}
\usepackage{color,diagbox}

\begin{document}
\title{Inhomogeneous chiral condensation under rotation in the holographic QCD}

\author{Yidian Chen}
\email[]{chenyidian@ucas.ac.cn} 
\affiliation{School of Nuclear Science and Technology, University of Chinese Academy of Sciences, Beijing 100049, China} 

\author{Danning Li}
\email[]{lidanning@jnu.edu.cn}
\affiliation{Department of Physics and Siyuan Laboratory, Jinan University, Guangzhou 510632, P.R. China} 

\author{Mei Huang}
\email[]{huangmei@ucas.ac.cn}
\affiliation{School of Nuclear Science and Technology, University of Chinese Academy of Sciences, Beijing 100049, China}	

\begin{abstract}
We investigate inhomogeneous chiral condensation under rotation considering finite size effects and boundary conditions in the holographic QCD model. The rotational suppression effect determined by $\Omega r$ is confirmed in the holographic model which is not influenced by the boundary conditions. For chiral condensation at the center, it is found that under Neumann boundary condition the finite size exhibits two opposite effects, i.e., catalysis at high temperatures and inverse catalysis at low temperatures. In contrast, under Dirichlet boundary condition, the effect of finite size on condensation is inverse catalysis, and small size induces a phase transition from inhomogeneous to homogeneous phase. The temperature-angular velocity phase diagrams of QCD are obtained for different boundary conditions and sizes, and it is found that the critical temperature decreases with angular velocity.
\end{abstract}
\maketitle

\section{Introduction}

The rotating matter is widespread in nature and can exhibit interesting properties and phenomena which are different from those of ordinary matter. 
For Bose-Einstein condensates or ultracold atom systems, the inhomogeneous lattice patterns of quantized vortices are formed in a rapidly rotating superfluid \citep{williams1974photographs,yarmchuk1979observation}.
In astrophysics, a spinning millisecond pulsar produces continuous gravitational waves, which can be detected by LIGO and Virgo \citep{Jaranowski:1998qm,LIGOScientific:2017csd,LIGOScientific:2022pjk}. In non-central heavy-ion collision (HIC) experiments, relativistic rotating quark-gluon plasmas (QGP) can be created with about $10^4-10^5\hbar$ angular momentum \citep{Muller:1983ed,Heinz:2000bk,STAR:2017ckg}, and 
the well-known chiral vortical effect \citep{Kharzeev:2007tn,Son:2009tf,Kharzeev:2010gr} and chiral vortical wave \citep{Jiang:2015cva} may occur in the chiral fluids due to anomalous transport phenomena induced by rotation.  

The QGP fireballs are produced by the HIC, which can be under extreme environments such as high temperature and/or density.  For the non-central collision case, the strong magnetic field and the high vorticity can be produced in the early stage of QGP, and the local angular velocity can reach 0.2 GeV with a relatively long lifetime according to the transport model \citep{Jiang:2016woz,Xia:2018tes,Wei:2018zfb} and hydrodynamic \citep{Becattini:2007sr} simulations. It is known that magnetic fields yield interesting effects on QCD phase diagram, e.g., magnetic catalysis \cite{Klevansky:1989vi,Klimenko:1990rh,Gusynin:1995nb} and inverse magnetic catalysis \cite{Bali:2011qj,Bali:2012zg,Bali:2013esa}. It is expected that analogous to magnetic fields, the rotation will also have some impact on the chiral condensation thus on the phase structure of QCD. As opposed to the magnetic field, the rotation not only affects the fermion pairing but also presses the matter toward the edge of the system. In analogy to the Tolman-Ehrenfest effect in the non-inertial frame, the condensation under rotation is considered inhomogeneous, for example, the formation of vortices is a general feature in rotating trapped Bose-Einstein condensates \citep{Madison:2000zz,Fetter:2009zz}.

For inhomogeneous chiral condensation formed by rotation, two factors are worth paying attention to: the spatial distribution of rotational speed and the finite size effect from the speed-of-light limit. Under the local density approximation, the angular velocity $\Omega$ of the system can be considered uniform, and in this case Refs.\citep{Chen:2015hfc,Jiang2016,Ebihara2017} studied radial coordinate $r$ dependent chiral condensation. These literatures pointed out that the rotation can be interpreted as an effective chemical potential, manifesting as a local suppression effect. Except near the boundary, the variation of condensation with radial direction is mild, which verifies the soundness of the approximation. In addition, Ref.\citep{Wang2019} identified another inhomogeneous effect, namely the centrifugal-like effect, which appears only in the nonuniform-$\Omega$ case and comes from the contribution of $\partial_r \Omega$.

The temperature-rotation $T-\Omega$ phase diagram of chiral condensation was explored as shown in Ref.\citep{Jiang2016}. Similar to the temperature-chemical potential phase diagram of QCD, the $T-\Omega$ diagram contains a new critical end point located at $T_{CEP}\simeq 0.020$GeV and $\Omega_{CEP}\simeq 0.644$GeV, which connects the crossover at high temperature and the first-order phase transition at low temperature. In Ref.\citep{Jiang2016}, the effective model with four-fermion interactions was used to describe the pairing phenomena and its corresponding Hamiltonian includes the angular velocity $\Omega$ and angular momentum $J$ coupling terms $\Omega\cdot J$. Not surprisingly, the acting of the angular velocity $\Omega$ and the chemical potential $\mu$ are pretty comparable, owing to the similarity of the terms $\Omega\cdot J$ and the chemical potential-conserved charge coupling term $\mu\cdot N$.

The finite size effect under rotation must be taken into account, and the results are sensitive to the choice of boundary conditions. The effect of finite size on phase transitions without rotation has been extensively studied \citep{Braun2005,Braun2006,Kiriyama2006,Shao2006,Palhares2011,Braun2012,Almasi2017,Klein2017,Wang2018,Li2019,Xu2020,Zhao2020}. As in Ref.\citep{Xu2020}, chiral condensation exhibits either catalysis or inverse catalysis for different boundary conditions and shows a quantized first-order phase transition in the case of periodic boundary condition. The boundary effects for the spatial profile of condensation under rotation have been discussed in Ref.\citep{Ebihara2017}, and it was found that the condensation shows as a plateau away from the boundary and oscillating near the boundary.

The non-perturbative characteristics of QCD in the infrared (IR) region give a great challenge in solving hadron physics and QCD phase structure. Various non-perturbative methods, such as lattice QCD \citep{Kogut:1982ds,Gupta:1997nd,Bloch:2003sk,Fodor:2012gf}, Dyson-Schwinger equations \citep{Alkofer:2000wg,Bashir:2012fs}, and functional renormalization group equations \citep{Wetterich:1992yh,Pawlowski:2005xe,Gies:2006wv}, are applied to solve this problem. At the end of the 20th century, the discovery of the anti-de Sitter/conformal field theory (AdS/CFT) correspondence or gauge/gravity duality \citep{Maldacena:1997re,Gubser:1998bc,Witten:1998qj} provided a new insight for handling non-perturbative problems. According to the holographic principle, a $D$-dimensional field theory is dual to the $(D+1)$-dimensional gravitational theory, and the extra dimensions can be understood as renormalization group flows \citep{Adams:2012th}. Starting from string theory, various top-down models, such as the $D_3-D_7$ model \citep{Erdmenger:2007cm}, the Witten-Sakai-Sugimoto model \citep{Sakai:2004cn,Sakai:2005yt}, and the STU model \citep{Behrndt:1998jd,Mas:2006dy,Pourhassan:2012mso} have been widely studied to detect the properties of strongly coupled theories. Another method known as bottom-up approach,  such as the hard-wall model \citep{Erlich2005}, the soft-wall model \citep{Karch2006}, the Gubser model \citep{Gubser:2008ny,Gubser:2008yx,DeWolfe:2010he,Grefa:2021qvt}, the improved holographic QCD model \citep{Gursoy:2007cb,Gursoy:2007er,Gursoy:2010fj}, the refined model \citep{Yang:2014bqa}, the Dudal model \citep{Dudal:2017max}, and the dynamical holographic QCD model (DhQCD) \citep{Li2013,Li2013a,Huang:2013qvz,Chen:2022goa}  have extensively studied hadron spectra, thermodynamic and transport properties of QCD matter, and QCD phase transitions.


There are usually three ways to introduce rotation effects in holography as follows. The first is to study it in the rotating Kerr black hole background as in Refs.\citep{McInnes:2014haa,Erices:2017izj,Mcinnes:2018xxz,McInnes:2018mwj,Arefeva:2020jvo}. The second is to do a coordinate transformation to a static black hole as in Refs.\citep{Awad:2002cz,Sheykhi:2010pya,BravoGaete:2017dso,Nadi:2019bqu}. The confinement/deconfinement phase transition of QCD under rotation is studied in this way in Refs.\citep{Chen2021,Braga:2022yfe}. The third one is to introduce rotation in the gauge field and solve it under the probe approximation, as in Refs.\citep{Domenech2010,Keranen2010,Keranen2010a,Dias:2013bwa,Li2020}. In this paper, the third scenario is chosen since the chiral phase transition of QCD under rotation is investigated.

The paper is organized as follows. First, the five-dimensional holographic model is introduced in Sec.\ref{sec:action}. Next, the profile of chiral condensation under rotation is investigated in Sec.\ref{sec:chi-rot}. Again, the effect of finite size on condensation is considered in Sec.\ref{sec:finite}. Then, the temperature-angular velocity phase diagram is obtained in Sec.\ref{sec:phase-diagram}. Finally, a short summary and discussion is presented in Sec.\ref{sec:sum}.

\section{5d setup}
\label{sec:action}

The soft wall AdS/QCD model~\cite{Karch2006} offers a good scenario to study the chiral phase transition, and it is quite convenient to extend the system to study the transition under rotation. 
This holographic model could successfully describe the chiral first-order phase transition, second-order phase transition, and crossover, e.g. as shown in Refs.~\citep{Chelabi2016,Chelabi2016a,Chen2020a}, by mapping a complex matrix-valued scalar field $X^{\alpha\beta}$ to the 4D operator $\langle\bar{q^\alpha}q^\alpha\rangle$, with $\alpha,\beta$ the flavor indexes. 
In the model, the phase transition is realized by adding nonlinear terms in the potential and introducing a proper profile of the dilaton field $\Phi$. 
Since the rotation destroys spatial homogeneity, in principle, the metric and fields should depend on the four-dimensional spacetime coordinates $x^{\mu}$. However, due to the complexity of the full numerical solution of general relativity, one might take the probe approximation, i.e. neglecting the back-reaction to the background metric and considering the rotational effect on the flavor sector only.  
Furthermore, we would assume that the rotating axis is fixed. Then, it is natural to take the cylindrical coordinate.

Taking the above consideration, we take the AdS-Schwarzschild metric as following
\begin{equation}
ds^{2}=\frac{L^{2}}{z^{2}}[-f(z)dt^{2}+\frac{dz^{2}}{f(z)}+dr^{2}+r^{2}d\theta^{2}+dx_{3}^{2}],
\end{equation}
where $L$ is the AdS radius , and $f(z)=1-(\frac{z}{z_{h}})^{4}$
is the blackening factor with the black hole horizon $z_{h}$. In the case of flat boundary, $L$ will be cancelled in the equation of motion, so we set it to be $1$. According to the holographic dictionary, the temperature of the 4D system is mapped to the
Hawking temperature of the background black hole, i.e. $T=\frac{1}{\pi z_{h}}$. Here we take the notation in which $z$ is the fifth dimension,  $x_{3}$ is the
direction of the angular velocity $\Omega$ and $r$ is the distance to the rotating axis. 
As a result of the rotation, the probe action involves a new gauge field $A_{M}$, which is dual to the non-zero current operator $\mathcal{O}^{\mu}=\langle\bar{q}\gamma^{\mu}q\rangle$ induced by the rotation, besides the complex scalar field $X$. 
Finally, the probe action is given as
\begin{equation}
S_{M}=-\int d^{5}x\sqrt{-g}e^{-\Phi(z)}\Bigg\{\text{{\rm Tr}}[(D^{M}X)^{\dagger}(D_{M}X)+V_{X}(|X|)]+\frac{1}{4}F_{MN}F^{MN}\Bigg\},\label{eq:action}
\end{equation}
where $g$ is the determinant of metric, $\Phi(z)$ is the dilaton
field, the covariant derivative is defined as $D_{M}X=\partial_{M}X-iA_{M}X$,
and the field strength is written as $F_{MN}=\partial_{M}A_{N}-\partial_{N}A_{M}$. In this work, we will focus on the scalar background and $U(1)$ current, so the $\text{SU}(2)$ gauge field in the original soft wall model has been neglected.

According to Refs.\citep{Chelabi2016,Chelabi2016a}, the chiral phase transition can be described by the following form of dilaton field
\begin{equation}
\Phi(z)=-\mu_{1}z^{2}+(\mu_{1}+\mu_{0})z^{2}\tanh(\mu_{2}z^{2}).
\end{equation}
Following Refs.\citep{Li2013,Li2013a,Chelabi2016,Chelabi2016a},The values of the three parameters $\mu_0, \mu_1, \mu_2$ would be taken as $\mu_{0}=(0.43{\rm GeV})^{2},\mu_{1}=(0.83{\rm GeV})^{2},\mu_{2}=(0.176{\rm GeV})^{2}$
, by fitting the Regge slopes of light mesons, the vacuum value of the condensate and the transition temperature of the chiral phase transition. 

The spontaneous breaking of the $SU(N_{f})_{L}\times SU(N_{f})_{R}$
symmetry to the $SU(N_{f})_{V}$ subgroup is realized by
the non-vanishing vacuum expectation value $X_{0}=\frac{\chi}{\sqrt{2N_{f}}}I_{N_{f}\times N_{f}}$
of the complex scalar field $X$, with number of flavors $N_{f}$
and identity matrix $I$. In this work, we would focus on the case with $N_f=2$. According to the AdS/CFT dictionary, the
UV asymptotic behavior of the scalar field $\chi$ has the form $m_{q}\zeta z+...+\frac{\sigma}{\zeta}z^{3}$
with quark mass $m_{q}$, quark condensation $\sigma$ and the normalization constant
$\zeta=\frac{\sqrt{3}}{2\pi}$ \citep{Cherman2009}. In the chiral
limit $m_{q}=0$, the potential term $V(\chi)$ of the scalar field
determines transition order of the phase transition, which could be first- or second-order.
Therefore, the following potential term $V(\chi)$ is considered in the paper
\begin{equation}
    V(\chi)\equiv{\rm Tr}[V_{X}(|X|)] =\frac{m_{5}^{2}}{2}\chi^{2}+\upsilon_{3}\chi^{3}+\upsilon_{4}\chi^{4},
    \label{eq:potential-1}
\end{equation}
and the phase transition is implemented by selecting the appropriate parameters. 
In the potential, the five-dimensional mass square
is $m_{5}^{2}=-3$, based on the holographic dictionary, while a non-zero
value of the cubic coupling $\upsilon_{3}$ in Eq.(\ref{eq:potential-1})
yields the first-order phase transition. Moreover, the quartic coupling
$\upsilon_{4}$ in Eq.(\ref{eq:potential-1}) 
determines the value
of the quark condensation at zero temperature. 
In the paper, as in Refs.\citep{Chelabi2016,Chelabi2016a}, two cases are chosen as $(\upsilon_{3},\upsilon_{4})=(0,8)$ and $(\upsilon_{3},\upsilon_{4})=(-3,8)$, where the first one generates a second-order phase transition and the last one gives a first-order phase transition in the chiral limit.

\begin{figure}
\includegraphics[width=0.4\textwidth]{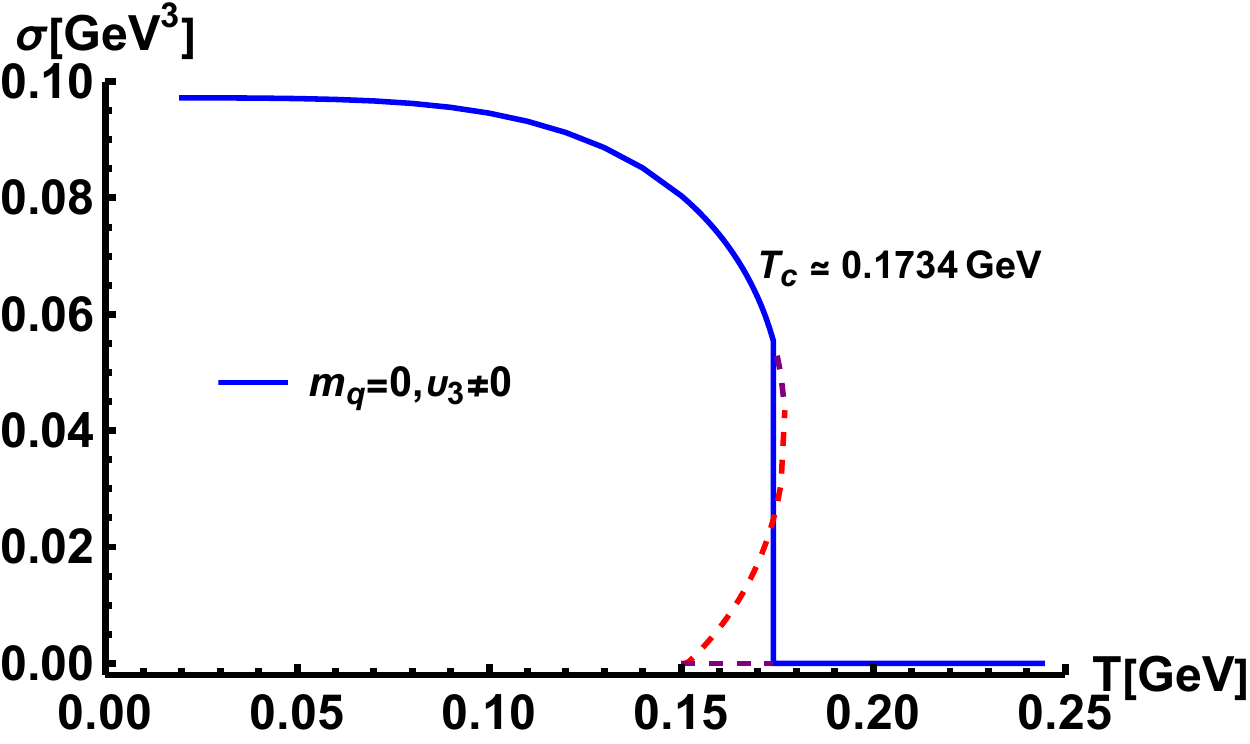}
\hspace*{1cm}
\includegraphics[width=0.4\textwidth]{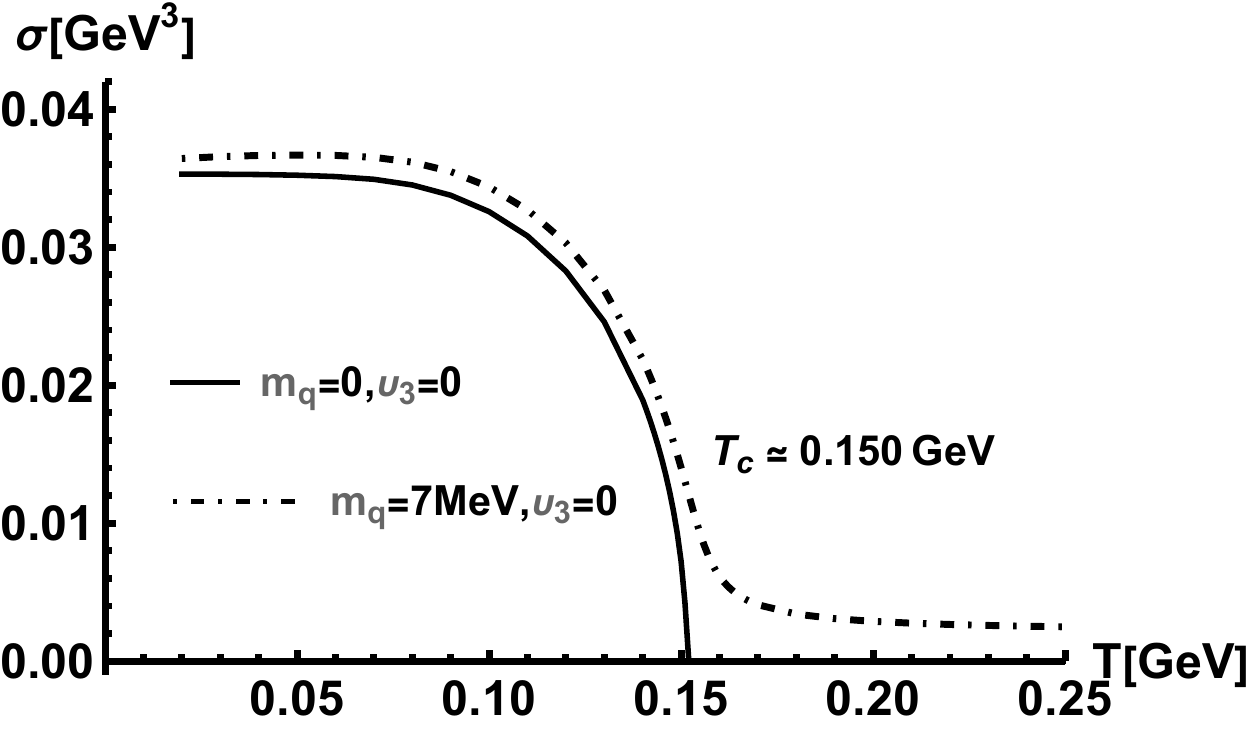}
\vskip -0.05cm \hskip 0 cm
\textbf{( a ) } \hskip 8 cm \textbf{( b )}
\caption{\label{fig:cpt-1} The chiral condensation as a
function of temperature $T$ with the scalar potential in Eq.(\ref{eq:potential-1}). In Panel.(a), the parameters are chosen as $(m_q,\upsilon_3,\upsilon_4)=(0,-3,8)$, where the blue solid line, red dashed line and purple dashed line represent the thermodynamic stable, metastable and unstable states of the first-order phase transition, respectively. The results of the second-order phase transition and crossover are shown in Panel.(b) with parameters of $(\upsilon_3,\upsilon_4)=(0,8)$, where the black solid and black dashed lines represent the quark masses are 0 and 7 MeV, respectively.}
\end{figure}

By taking proper values of the model parameters and solving the equation of motion, one could obtain the chiral condensate as a function of temperature. The Panel.(a) and Panel.(b) of Fig.\ref{fig:cpt-1} show the results for $(\upsilon_{3},\upsilon_{4})=(-3,8)$ and $(\upsilon_{3},\upsilon_{4})=(0,8)$, respectively.
In Fig.\ref{fig:cpt-1}(a),
the purple dashed line indicates the metastable solution and the red
dashed line denotes the unstable solution. 
In Fig.\ref{fig:cpt-1}(b), the difference between the black solid and dashed lines is the selection of the quark masses, which are 0 and 7 MeV, respectively.
It can be shown that the critical temperature $T_c\simeq 173.4$ MeV for the first-order phase transition, and $T_c\simeq 150$ MeV for the second-order phase transition and crossover.
For Fig.\ref{fig:cpt-1},
they are specifically described in Ref.\citep{Chelabi2016,Chelabi2016a} and will not be further clarified here. 

Considering a rotating system in which the direction of angular velocity $\Omega$ is fixed, both the scalar field 
$\chi$ and the gauge field $A_{M}$ depend on the fifth dimensional coordinate $z$ and radial coordinate $r$, 
yet independent of the polar coordinate $\theta$ and height
coordinate $x_{3}$, i.e.
\begin{equation}
\chi=\chi(z,r),\qquad\qquad A_{M}=A_{M}(z,r).
\end{equation}
In the $A_{z}=0$ gauge, the only nonzero component of the gauge field is $A_{\theta}$, which is dual to the polar current operator $\langle\bar{q}\gamma^\theta q\rangle$. To maintain the conservation of the vector current $j^\mu=\bar{q}\gamma^\mu q$ of the system, $\bar{q}\gamma^r q=0$ and $\bar{q}\gamma^{x_3} q=0$ are chosen, i.e., the system is in the steady state and no radial and vertical flows exist. And the non-zero polar direction component
$A_{\theta}$ is an effective polarization term $\overrightarrow{\Omega}\cdot\overrightarrow{J}$
in the dual field theory with angular momentum $\overrightarrow{J}$.
Therefore, the equations of motion of the action Eq.(\ref{eq:action})
are
\begin{align}
\partial_{z}(f\partial_{z}\chi)-f(\frac{1}{z}+\Phi^{\prime})\partial_{z}\chi+\partial_{r}^{2}\chi+\frac{\partial_{r}\chi}{r}+(\frac{3}{z^{2}}+\frac{f'}{z}-\frac{3f}{z^{2}}-\frac{f\Phi^{\prime}}{z}-\frac{3\upsilon_{3}\chi}{z}-4\upsilon_{4}\chi^{2}-\frac{A_{\theta}^{2}}{r^{2}})\chi & =0\label{eq:eom1-a}\\
\partial_{z}(f\partial_{z}A_{\theta})-f(\frac{1}{z}+\Phi^{\prime})\partial_{z}A_{\theta}+\partial_{r}^{2}A_{\theta}-\frac{\partial_{r}A_{\theta}}{r}-\chi^{2}A_{\theta} & =0\label{eq:eom2}
\end{align}
where $\chi$ is replaced by $z\chi$.

As in Refs.\citep{Domenech2010,Keranen2010,Keranen2010a,Li2020},
the appropriate boundary conditions are chosen in the following form
to solve the equations of motion. At the AdS conformal boundary, the
boundary conditions of the $\chi$ and $A_{\theta}$ fields are
\begin{equation}
\chi|_{z=0}=m_{q}\zeta,\qquad\qquad A_{\theta}|_{z=0}=\Omega(r)r^{2},
\end{equation}
where the angular velocity $\Omega(r)$ is not necessarily constant,
but spatially dependent. In general, $\Omega(r)$ might be solved in a full back-reaction model when choosing proper initial conditions. However, since we are working in probe limit due to the complexity of getting the full solution, we will take certain phenomenological configurations of $\Omega(r)$ instead. 

In this paper, three possible cases of angular
velocity configurations $\Omega(r)$ are investigated: (i) $\Omega=$constant;
(ii) $\Omega(r)=\omega_{0}({\rm exp}[1.5(r-r_{0})^{2}]+1)^{-1}$;
(iii) $\Omega(r)=\omega_{0}({\rm exp}[(r-r_{0})]+1)^{-1}$. The cases
(ii) and (iii) are from Ref.\citep{Wang2019}, where $\omega_{0}$
and $r_{0}$ are parameters. 
Among them, case (i) considers the rigid rotation of the system with a constant angular velocity; case (ii) is a vortex-like angular velocity distribution; case (iii) is an angular velocity distribution concentrated at the center.
For the sake of causality, the polar
direction is restricted to a finite size, i.e., the radius $R$ of
the system. Therefore, at the edges of the system, the boundary conditions
are taken as
\begin{equation}
\partial_{r}\chi|_{r=R}=0\quad{\rm (Neumann)}\quad{\rm or}\quad\chi|_{r=R}=0 \quad{\rm (Dirichlet)},\qquad\qquad A_{\theta}|_{r=R}=\Omega(R)R^{2},\label{eq:bcR}
\end{equation}
where $\chi$ is regarded as two possible boundary conditions, i.e.
Neumann or Dirichlet boundary condition. 
The Neumann boundary condition is chosen such that the order parameter is independent of $r$ in the region away from the angular velocity distribution. If the system is bounded in a box, the Dirichlet boundary condition can be taken into account.
At the center of the polar
direction, the following boundary conditions are considered
\begin{equation}
\partial_{r}\chi|_{r=0}=0,\qquad\qquad\partial_{r}A_{\theta}|_{r=0}=0,
\end{equation}
which guarantee the smoothness of the field configuration at the center.
With the above equations of motion and boundary conditions, the profile
of the chiral condensation under rotation can be obtained. 

By comparing the free energy of the inhomogeneous and homogeneous phases, the steady state of the system can be determined. According to the holographic principle, the partition functions of the gravitation and the boundary field theory are equivalent $\mathcal{Z}_{QCD}=\mathcal{Z}_{gravity}$, so the free energy of the system can be extracted from the on-shell action
\begin{eqnarray}
\mathcal{F}&=&\int_0^Rdr\int_0^{z_h}dz\sqrt{-g}e^{-\Phi}(-\frac{1}{2}\upsilon_3\chi^3-\upsilon_4\chi^4+\frac{1}{4}F^2)\nonumber\\
&&-\int_0^{z_h}dz\frac{1}{2}(\sqrt{-g}e^{-\Phi}\chi\partial^r\chi)\Big|_0^R-\int_0^Rdr\frac{1}{2}(\sqrt{-g}e^{-\Phi}\chi\partial^z\chi\Big|_0^{z_h}).
\end{eqnarray}

\section{chiral condensation under rotation\label{sec:chi-rot}}

In this section, the spatial profile of the chiral condensation
under rotation is discussed. As mentioned in the previous section,
due to the rotation of the system, nonzero polar orientation current
$\langle\bar{q}\gamma^{\text{\ensuremath{\theta}}}q\rangle$ is induced,
which further leads to spatially dependent chiral condensation. It
is worth stating that different spatial distribution of angular
velocity $\Omega(r)$ will lead to different configurations of chiral condensation.
Therefore, the three cases (i), (ii), and (iii) introduced before
are considered, i.e., uniform angular velocity, vortex-like distribution
of rotational speed and centered distribution.
Of course, the selection of parameters for cases 
(ii) $\Omega(r)=\omega_{0}({\rm exp}[1.5(r-r_{0})^{2}]+1)^{-1}$ and 
(iii) $\Omega(r)=\omega_{0}({\rm exp}[(r-r_{0})]+1)^{-1}$ need
to satisfy the causality constraint. Again as a consequence of the
causality requirement, the system is restricted to the radius $R$.
Hence, in this section, the radius is fixed to the typical quark gluon
plasma scale 4 fm, i.e., 20${\rm GeV}^{-1}$, so as to avoid the finite
size effect, which will be discussed in the next section. Furthermore,
two possible boundary conditions for the scalar field $\chi$ as stated
in Eq.(\ref{eq:bcR}) are illustrated in Sec.\ref{subsec:rot-bc-a}
and Sec.\ref{subsec:rot-bc-b}, respectively.


\subsection{Neumann boundary condition\label{subsec:rot-bc-a}}

In this subsection, chiral condensation as a function of radial coordinate $r$ is studied. Since it is insignificant whether the phase transition
is first-order, second-order or crossover on the qualitative results,
the example is illustrated in the form of the potential term Eq.(\ref{eq:potential-1}) with the parameter $(m_{q},\upsilon_{3},\upsilon_{4})=(0,-3,8)$
chosen. 

\begin{figure}
\includegraphics[width=0.38\textwidth]{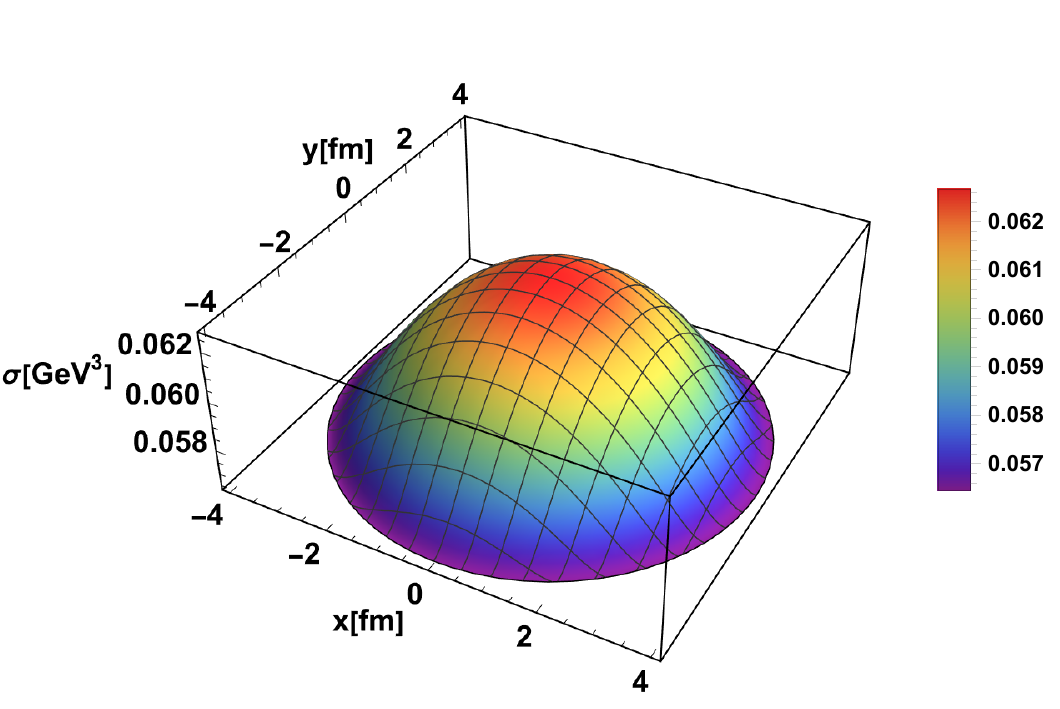}
\hspace*{1cm}
\includegraphics[width=0.45\textwidth]{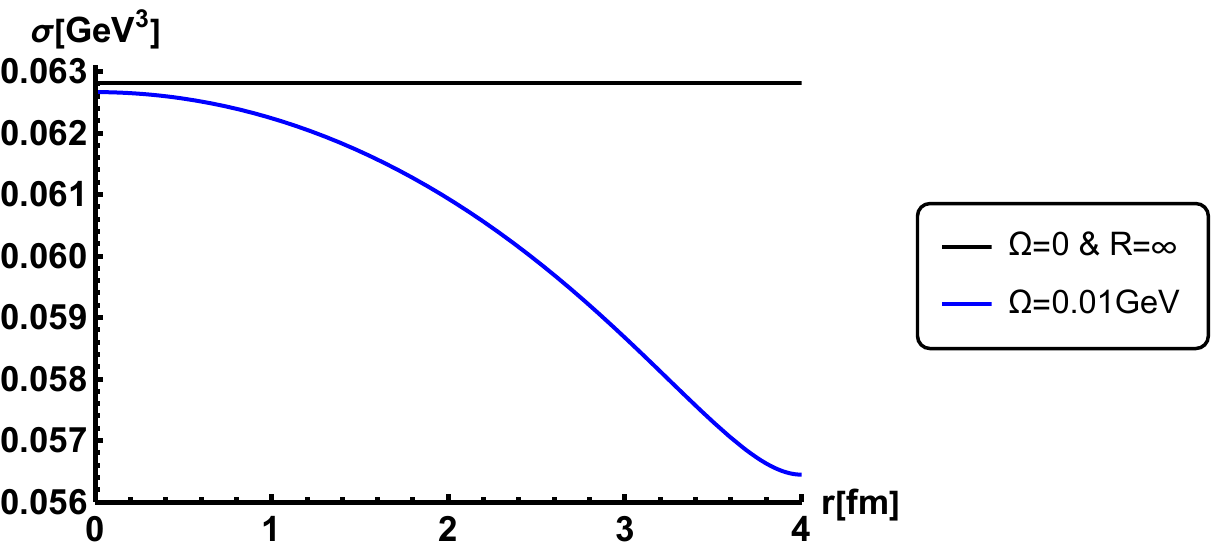}
\vskip -0.05cm \hskip -2 cm
\textbf{( a ) } \hskip 7 cm \textbf{( b )}
\caption{\label{fig:sigmar_N} 3D and 2D plots of chiral condensation as a function of radial $r$ at $T= 170 $MeV and $\Omega= 0.01$ GeV with NBC and $(m_q,\upsilon_{3},\upsilon_{4})=(0,-3,8)$. In Fig.(b),  the black line indicates the value of condensation at the same temperature without rotation and finite size.}
\end{figure}

For the temperature $T=170$ MeV and the uniform angular velocity $\Omega=0.01$ GeV, the configuration of the condensation $\sigma(r)$ in space
is shown in Fig.\ref{fig:sigmar_N}. Among them, the left panel is
a three-dimensional plot, while the right two-dimensional plot with
the horizontal coordinate $r$ from 0 to 4 fm. As can
be seen in the figure, the condensation is largest at the center and
decreases with increasing radial coordinate $r$, similar to a swelling. 
The condensation has a maximum value of $0.0626~{\rm GeV}^3\simeq (0.397~{\rm GeV})^3$ at the center, and it gradually decreases to $0.0564~{\rm GeV}^3\simeq (0.383~{\rm GeV})^3$ with increasing $r$.
It is worth noting that even at the center, the value of the condensation
does not reach the non-rotational value $0.0628~{\rm GeV}^3\simeq (0.398~{\rm GeV})^3$. It arises from the combined
effect of rotation and finite size of the system. For the effect of
finite size, it will be explained in the next section. Furthermore,
the concrete shape of the condensation depends on the form of the angular
velocity function and the boundary conditions of the scalar field
at the edges. The Neumann boundary condition (NBC) is chosen here, while
the Dirichlet boundary condition (DBC) will be shown in the next subsection.

\begin{figure}
\includegraphics[width=0.45\textwidth]{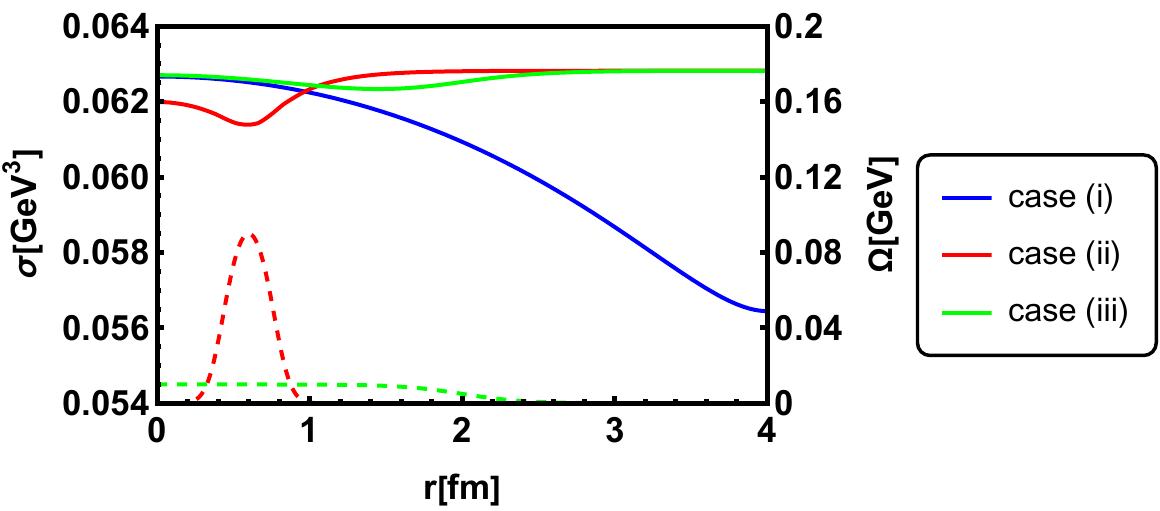}
\hspace*{1cm}
\includegraphics[width=0.45\textwidth]{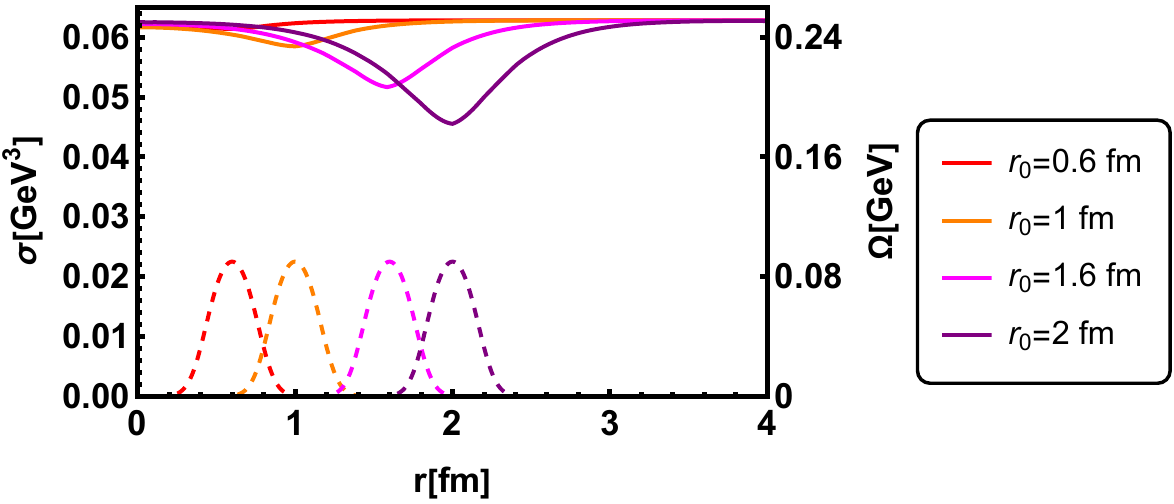}
\vskip -0.05cm \hskip -1.8 cm
\textbf{( a ) } \hskip 8.3 cm \textbf{( b )}
\caption{\label{fig:sigmacompare_N} The chiral condensation as a function of radial $r$ at $T = 170$ MeV with NBC and $(m_q,\upsilon_{3},\upsilon_{4})=(0,-3,8)$, where the solid and dashed lines denote the profile of the condensation and the distribution of the angular velocity, respectively. In (a), the three cases of angular velocity distribution are: (i) $\Omega=0.01$, (ii) $\Omega(r)=0.18({\rm exp}[1.5(r-10)^{2}]+1)^{-1}$ and (iii) $\Omega(r)=0.01({\rm exp}[(r-10)]+1)^{-1}$. Fig.(b) represents case (ii) $\Omega(r)=0.18({\rm exp}[1.5(r-r_0)^{2}]+1)^{-1}$ with $r_0=0.6$ fm, 1 fm, 1.6 fm and 2 fm.}
\end{figure}

The deformation of condensation by different distributions of angular velocity
is shown in Fig.\ref{fig:sigmacompare_N}(a) with the case (ii) $\Omega(r)=0.18({\rm exp}[1.5(r-3)^{2}]+1)^{-1}$ and case (iii) $\Omega(r)=0.01({\rm exp}[(r-10)]+1)^{-1}$. The dashed and solid lines
in the figure indicate the variation of angular velocity and condensation
with radial coordinate $r$, respectively. From the red line, it can be seen that
the angular velocity takes a maximum value at radial $r=0.6$ fm, meanwhile
the condensation obtains a minimum value. At the center and the
edge $R$, the angular velocity tends to 0 and the condensation goes to a stable value. 
The results suggest that for the case (ii), the bigger the angular velocity, the more noticeable
the suppression of condensation, in agreement with Ref.\citep{Wang2019}.
This conclusion can be seen from $\frac{A_\theta^2}{r^2}$ in Eq.\eqref{eq:eom1-a}. As the angular velocity increases, the term $\frac{A_\theta^2}{r^2}$ becomes more crucial, which is equivalent to increasing the five-dimensional mass $m_5^2$ and thus reducing the chiral condensation. In Ref.\citep{Wang2019}, it pointed out that it is $\Omega r$ instead of the angular velocity $\Omega$ that determines the suppression effect, and this conclusion is also reproduced in the holographic model. As shown in Fig.\ref{fig:sigmacompare_N}(b), the condensation profile for case (ii) with different parameters $r_0$ are exhibited. It can be seen that although the distribution of the angular velocity is similar except for shifting the center, the condensation decreases significantly as the radial $r$ becomes larger. It can be further seen from Eq.\eqref{eq:eom1-a} that the term $\frac{A_\theta^2}{r^2}$ is converted to $\Omega^2r^2$ when $A_\theta\simeq \Omega r^2$ is considered, then it is $\Omega r$ rather than the angular velocity $\Omega$ that increases the five-dimensional mass.

For the green line, the shape of the condensation with radial coordinate $r$ is
similar to that of the red line, both getting a minimum at a particular
position and reaching a stable value at the center and the edges. 
The difference is that the angular velocity in case (iii) is approximately non-zero constant near the
center, so the suppression effect is similar to case (i). 
However, near the edges, the green line converges to the red line. It should be noted that
this result of case (iii) is not consistent with Ref.\citep{Wang2019}, where the enhancement of condensation by the first derivative of the angular velocity is not exhibited in the holographic model.

\begin{figure}
\includegraphics[width=0.45\textwidth]{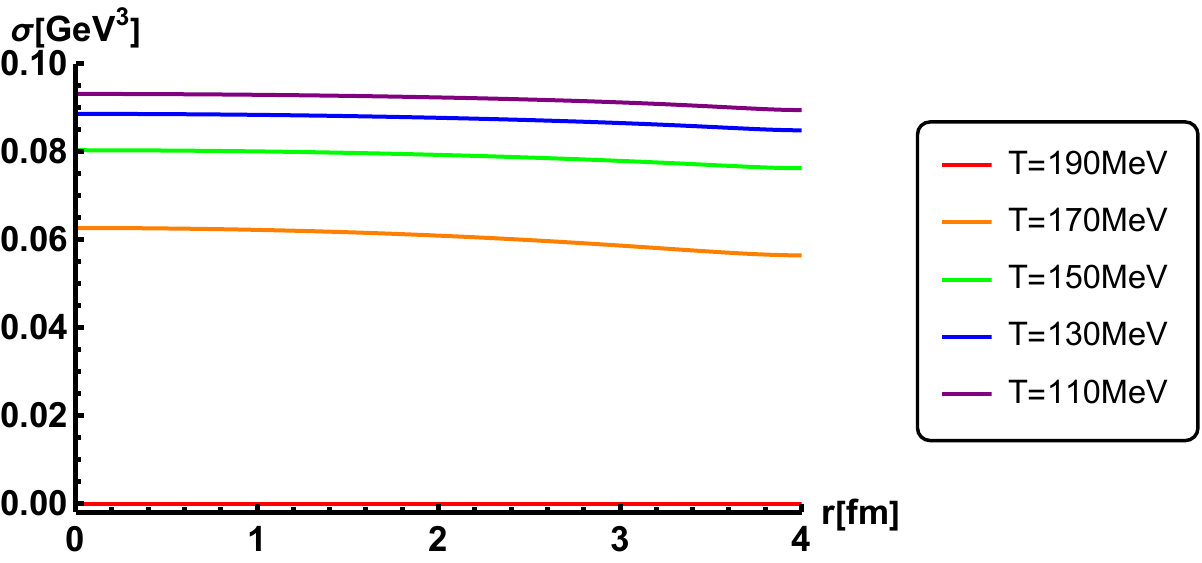}
\hspace*{1cm}
\includegraphics[width=0.45\textwidth]{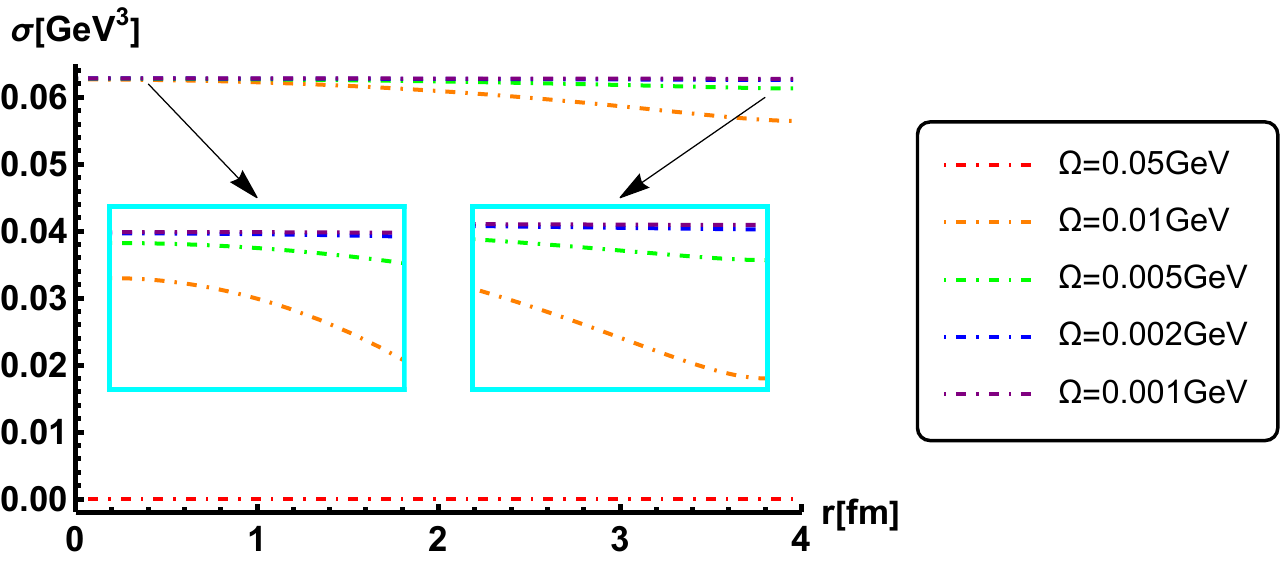}
\vskip -0.05cm \hskip -2.2 cm
\textbf{( a ) } \hskip 8.1 cm \textbf{( b )}
\caption{\label{fig:sigmaT_N} The chiral condensation as a function of radial coordinate $r$ with NBC and $(m_q,\upsilon_{3},\upsilon_{4})=(0,-3,8)$. Here, the solid and dashed lines indicate the fixed angular velocity $\Omega=0.01$GeV and temperature $T=170$MeV, respectively.}
\end{figure}

In the following, returning to the case of uniform angular velocity,
the role of different temperatures and angular velocities on condensation
is investigated. The shape of condensation with radial $r$ is taken
in Fig.\ref{fig:sigmaT_N}(a) for fixing the angular velocity to 0.01
GeV and varying the temperature to 110, 130, 150, 170 and 190 MeV. 
It can be seen from the figure that the shape of the condensation tends to be the same at different temperatures and that the change in condensation is small as $r$ increases. The value at the center increases from $0.0626~{\rm GeV}^3\simeq (0.397~{\rm GeV})^3$ to $0.094~{\rm GeV}^3\simeq (0.454~{\rm GeV})^3$ as the temperature decreases from 170 MeV to 110 MeV.
And for the red line in the figure, i.e., when the temperature reaches 190 MeV, the system enters into the homogeneous chiral symmetric phase. Consequently, it can
be expected that as the temperature increases, there exists a certain
critical temperature $T_c$, across which the system will enter the homogeneous chiral
restoration state and the condensation will no longer depend on space.
Further, at least for first- and second-order phase transitions, the
phase transition can be defined in this inhomogeneous phase, i.e.,
when the condensation at the full space or center enters into chiral
restoration, and thus the $T-\Omega$ phase diagram of the chiral
condensation can be drawn.

Here, fixing the temperature at 170 MeV and varying the angular velocity
at 0.001, 0.002, 0.005, 0.01, and 0.05 GeV, the variation of the shape of the condensation
with the angular velocity is explored and the results are shown in
Fig.\ref{fig:sigmaT_N}(b). 
It can be obtained that the increasing angular velocity suppresses the condensation, especially at the edges. As the angular velocity increases from 0.001 GeV to 0.01 GeV, the change at the center is negligible and the condensation decreases from $0.0626~{\rm GeV}^3\simeq (0.397~{\rm GeV})^3$ to $0.0564~{\rm GeV}^3\simeq (0.383~{\rm GeV})^3$ at the edges. The greater angular velocity means that greater deformation for condensation.
Similar to the chemical potential, the angular velocity plays the inverse catalysis effect.
It is worth remarking here that simply reducing the angular
velocity does not allow the condensation value to reach the no-rotation
case, because the finite size effect also remains. And for the red
line in the figure, i.e., when the angular velocity reaches 0.05 GeV,
the system enters the homogeneous chiral restoration phase and the value of condensation
in space is 0, which is analogous to the previous assertion.

\subsection{Dirichlet boundary condition\label{subsec:rot-bc-b}}

In this subsection, chiral condensation as a function of radial $r$
is discussed when the scalar field at the edge takes the DBC. Similar to the previous subsection, The potential term Eq.(\ref{eq:potential-1}) with the parameters $(m_{q},\upsilon_{3},\upsilon_{4})=(0,-3,8)$ is selected as an example for illustration. 

\begin{figure}
\includegraphics[width=0.38\textwidth]{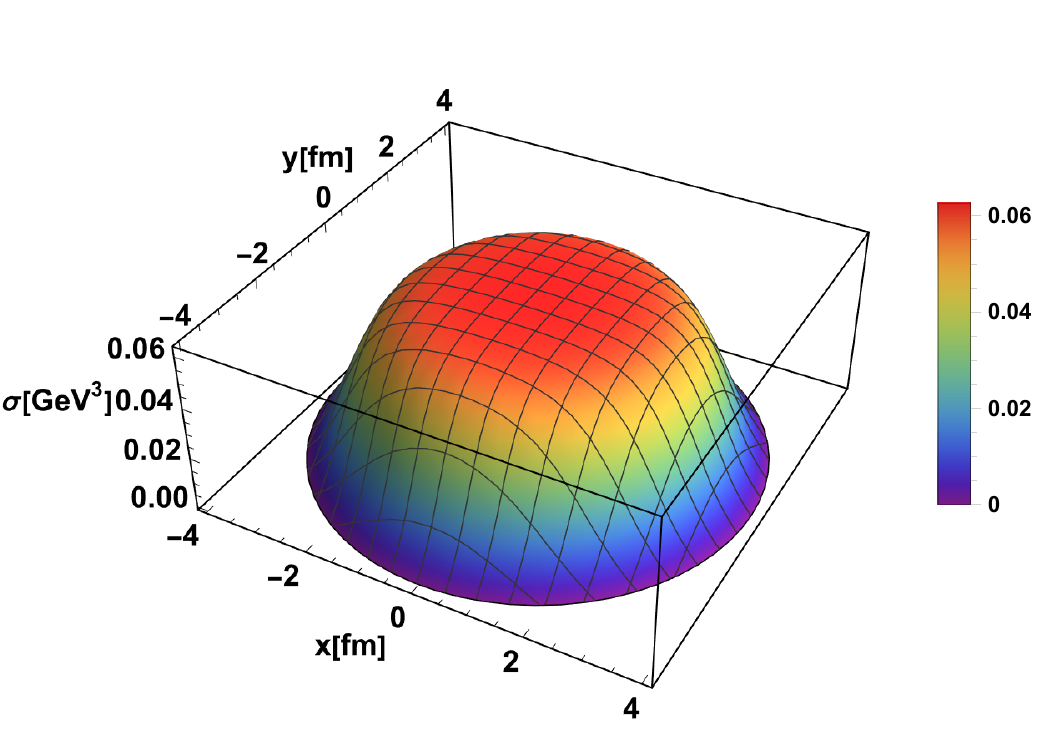}
\hspace*{1cm}
\includegraphics[width=0.45\textwidth]{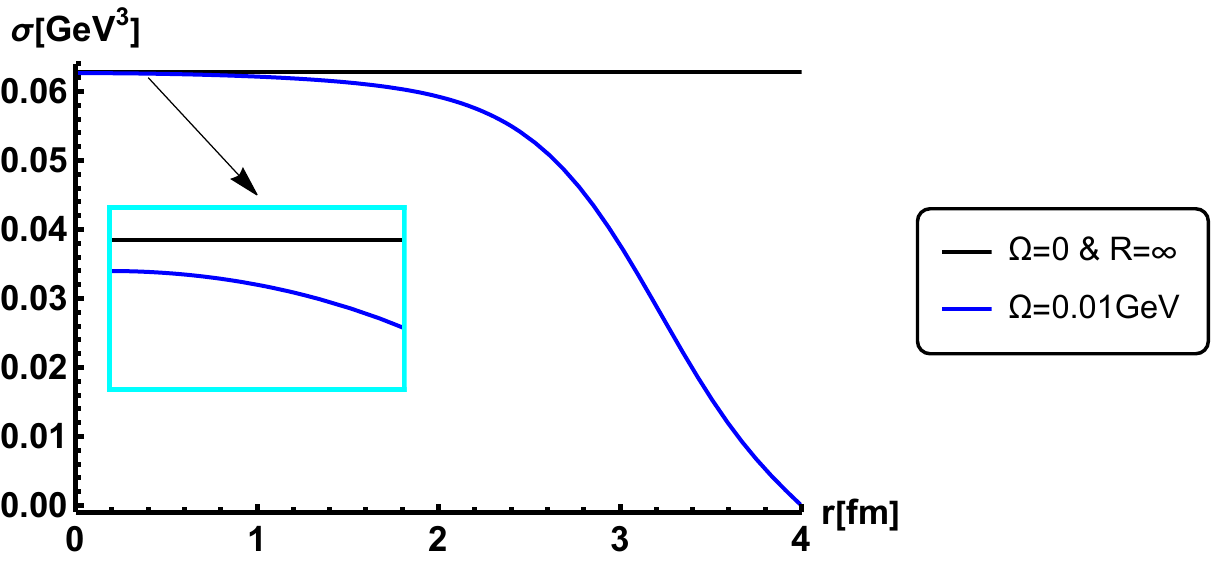}
\vskip -0.05cm \hskip -2 cm
\textbf{( a ) } \hskip 7 cm \textbf{( b )}
\caption{\label{fig:sigmar_D} 3D and 2D plots of chiral condensation as a function of radius $r$ at $T= 170 $MeV and $\Omega= 0.01$ GeV with DBC and $(m_q,\upsilon_{3},\upsilon_{4})=(0,-3,8)$. In Fig.(b), 
the black line indicates the value of condensation at the same temperature without rotation and finite size.}
\end{figure}

For a temperature of 170 MeV and an angular velocity of 0.01 GeV,
the 3D and 2D plots of the spatial profile of the condensation
are shown in Fig.\ref{fig:sigmar_D}. From the figure, it can be observed
that the value of condensation changes very slowly at radial coordinate less than
2 fm, while at radial coordinate greater than 2 fm,
it decreases rapidly until 0. The shape of the condensation on the
space is similar to a drum, which is distinct from the NBC. 
At the center, the value of condensation is about $0.0625~{\rm GeV}^3\simeq (0.396~{\rm GeV})^3$, which is comparable to the results of NBC.
Again, this phenomenon arises from the
finite size effect of the system. Therefore, for angular velocity
is not so large, the value of condensation near the center does not
depend significantly on the boundary conditions at the edge. The plateau profile of condensation near the center is consistent with Ref.\citep{Ebihara2017}, further demonstrating the validity of the local density approximation, while holography does not find oscillating behavior at the edges $R$.

\begin{figure}
\includegraphics[width=0.45\textwidth]{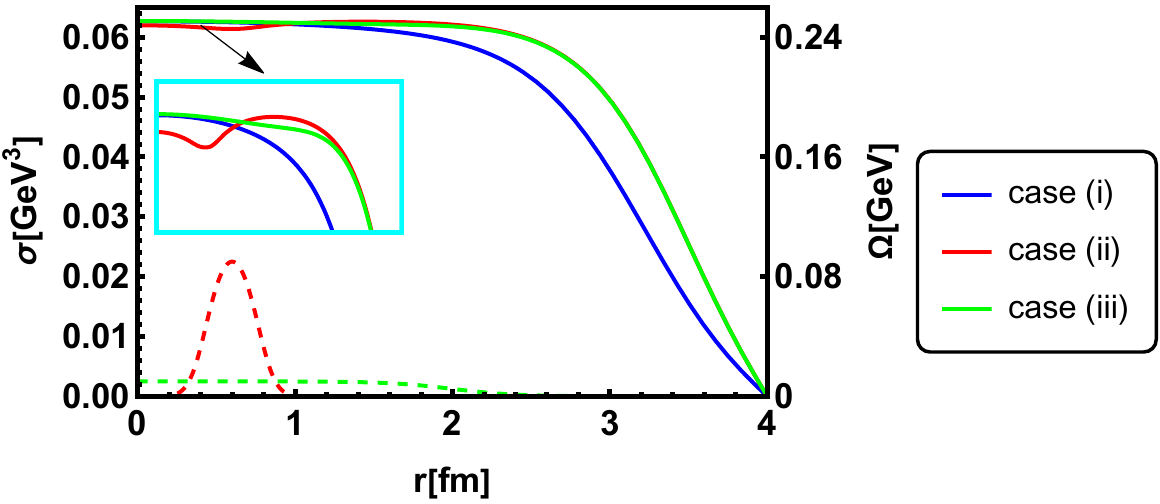}
\hspace*{1cm}
\includegraphics[width=0.45\textwidth]{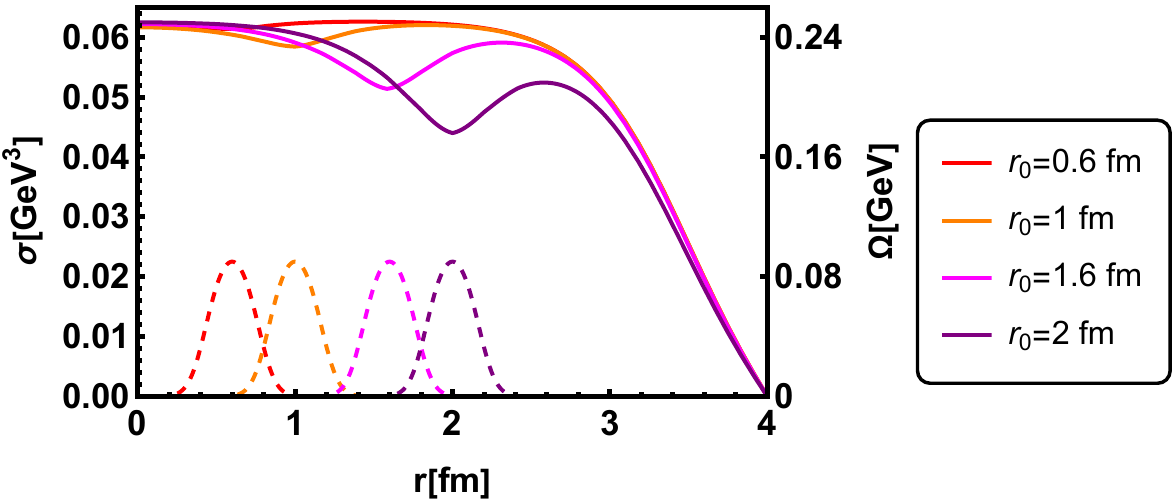}
\vskip -0.05cm \hskip -1.8 cm
\textbf{( a ) } \hskip 8.3 cm \textbf{( b )}
\caption{\label{fig:sigmacompare_D} The figure shows the chiral condensation as a function of radial $r$ at $T = 170$ MeV with DBC and $(m_q,\upsilon_{3},\upsilon_{4})=(0,-3,8)$, where the solid and dashed lines denote the profile of the condensation and the distribution of the angular velocity, respectively. In Fig.(a), the three cases of angular velocity distribution are: (i) $\Omega=0.01$, (ii) $\Omega(r)=0.18({\rm exp}[1.5(r-10)^{2}]+1)^{-1}$ and (iii) $\Omega(r)=0.01({\rm exp}[(r-10)]+1)^{-1}$. Fig.(b) represents case (ii) $\Omega(r)=0.18({\rm exp}[1.5(r-r_0)^{2}]+1)^{-1}$ with $r_0=0.6$ fm, 1 fm, 1.6 fm and 2 fm.}
\end{figure}

The deformation of the angular velocity function on the condensation
when the parameters of cases (ii) and (iii) are chosen as in the previous
subsection is shown in Fig.\ref{fig:sigmacompare_D}. For the red
line, the condensation yields a local minimum at the maximum angular
velocity, but it tends to 0 near the edge $R$. This is
not essentially different from the conclusion of the NBC, and the behavior at the edges is driven by the boundary condition.
As can be seen in Fig.\ref{fig:sigmacompare_D}(b), the suppressive effect of $\Omega r$ is also seen in DBC, consistent with NBC.
For the green line, its behavior near the center is similar to that of the blue line, while near the edge is similar to that of the red line. 
It is worth mentioning that the centrifugal-like effect of condensation does not appear in the holographic model for either Neumann or Dirichlet boundary conditions, inconsistent with Ref.\citep{Wang2019}.
The possible reason for this is that the introduction of rotation
in the holographic model comes from the gauge field, which acts analogous
to the chemical potential and therefore causes the system to tend
to chiral restoration, reducing the value of condensation.

\begin{figure}
\includegraphics[width=0.45\textwidth]{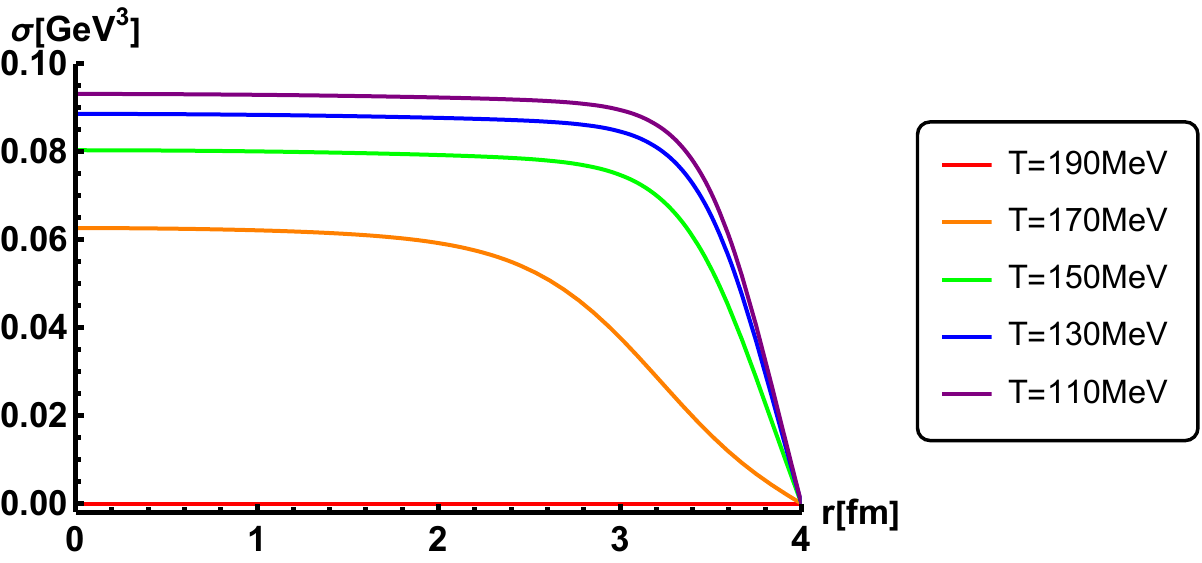}
\hspace*{1cm}
\includegraphics[width=0.45\textwidth]{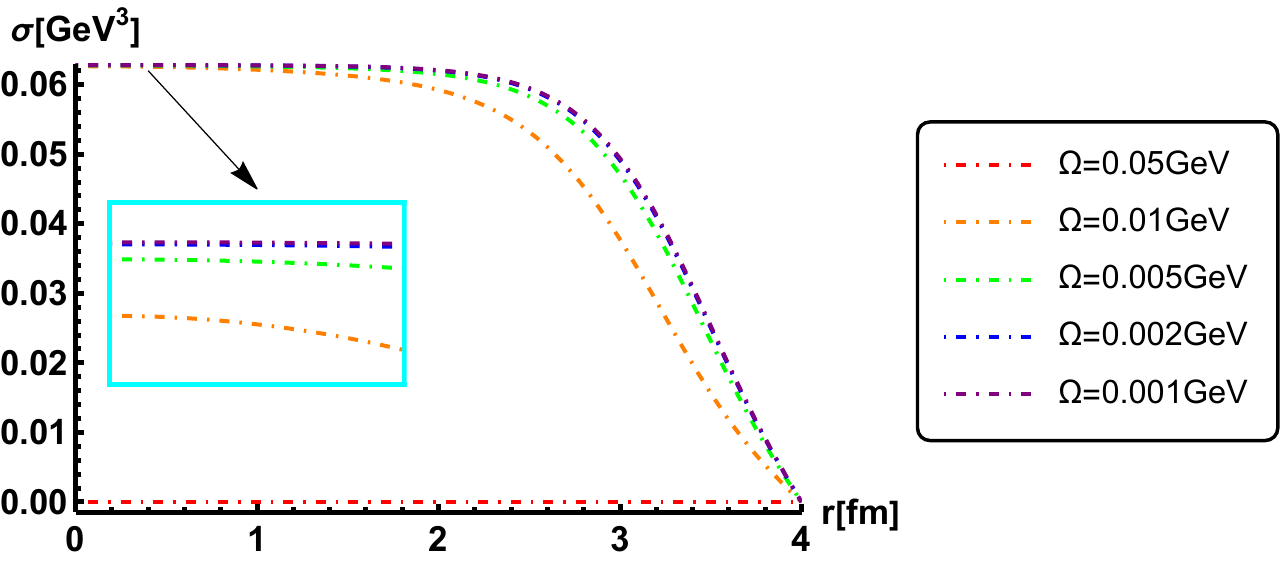}
\vskip -0.05cm \hskip -2.2 cm
\textbf{( a ) } \hskip 8.1 cm \textbf{( b )}
\caption{\label{fig:sigmaT_D} The chiral condensation as a function of radial coordinate $r$ with DBC and $(m_q,\upsilon_{3},\upsilon_{4})=(0,-3,8)$. Here, the solid and dashed lines indicate the fixed angular velocity $\Omega=0.01$GeV and temperature $T=170$MeV, respectively.}
\end{figure}

The condensation functions for different temperatures at fixed angular
velocity $\Omega=0.01$ GeV are displayed in Fig.\ref{fig:sigmaT_D}(a). 
As we can see in the figure, the value at the center increases from $0.0625~{\rm GeV}^3\simeq (0.396~{\rm GeV})^3$ to $0.093~{\rm GeV}^3\simeq (0.453~{\rm GeV})^3$ as the temperature decreases from 170 MeV to 110 MeV.
The special feature is that
the \textquotedbl surface\textquotedbl{} of the \textquotedbl drum\textquotedbl{}
condensation shrinks as the temperature increases. The plateau surface
is about 1.5 fm for a temperature of 170 MeV, while the
surface expands to 3 fm for a temperature of 130 MeV. 
Therefore, it can be expected that at 0 temperature, 
the condensation is approximately \textquotedbl homogeneous\textquotedbl{} except near the edges.
In turn, it can be predicted that, similar to NBCs,
DBCs also have phase transitions, and their
effects on the $T-\Omega$ phase diagram will be mentioned in Sec.\ref{sec:phase-diagram}.

When the temperature is fixed at 170 MeV, condensation with different
angular velocities is shown in Fig.\ref{fig:sigmaT_D}(b). 
Similarly, the angular velocity does not change much for condensation at the center, but the variation can reach $0.01~{\rm GeV}^3$ near the edges.
It is noteworthy
that at an angular velocity of 0.05 GeV, the system recovers to a
state of chiral symmetry, which is similar to the NBC.

\section{finite size effect}
\label{sec:finite}

In the previous section, the chiral condensation under rotation, including
various angular velocities and temperatures, was discussed. There
the size of the system was fixed to $R=4{\rm fm}\simeq20{\rm GeV}^{-1}$, which is comparable to the size of the fireball created in the current experiments.
As mentioned above, in a rotating system, the size effect might be important. Also, the exact size of the fireball is hard to be extracted. Thus, in this section, we will study the finite-size effect on condensation. It is common knowledge that the thermal and dense
quark matter produced in heavy ion collisions is bounded in a finite
volume at the nuclear scale. Consequently, it is highly essential
to recognize the effect of finite size on the QCD matter and phase
structure, which has been extensively studied during
the past decades\citep{Braun2005,Braun2006,Kiriyama2006,Shao2006,Palhares2011,Braun2012,Almasi2017,Klein2017,Wang2018,Li2019,Xu2020,Zhao2020}.
As shown in the research of Ref.\citep{Xu2020}, under suitable boundary
conditions, the catalysis or inverse catalysis of chiral symmetry
breaking occurs as the size of the system decreases. Of course, these
studies did not consider the inhomogeneous chiral condensation on
the spatial coordinate $x$. Therefore, the effect of finite size
on the spatially dependent condensation, induced by the rotation of
the system, would be an interesting issue.

The profile of chiral condensation under rotation has been obtained
in the holographic model. The results show that the effect of rotation,
similar to the chemical potential, is inverse catalysis,
regardless of NBC or DBC, while the boundary conditions only affect
the condensation near the edges. Nevertheless, when finite size effects are included, 
different boundary conditions will produce different results, 
which are also comparable to the NJL model.
In the NJL model, as in Ref.\citep{Xu2020}, periodic and antiperiodic
boundary conditions enhance symmetry breaking and restore chiral symmetry,
respectively, as the system size decreases. Therefore, in the following,
NBC and DBC will be chosen to evaluate the effect of finite size on the
chiral condensation and phase structure.

\subsection{Catalysis and inverse catalysis}

In this subsection, the effect of finite size on the condensation
is investigated, considering NBC at the edges. Here, the five different radius 
of the system are taken into account as 1-5 fm, i.e., 5-25
${\rm GeV}^{-1}$, and marked with the colors red, orange, green,
blue and purple, respectively. As in the previous section, the potential Eq.(\ref{eq:potential-1}) with $(m_{q},\upsilon_{3},\upsilon_{4})=(0,-3,8)$ is exhibited as an example.

\begin{figure}
\includegraphics[width=0.36\textwidth]{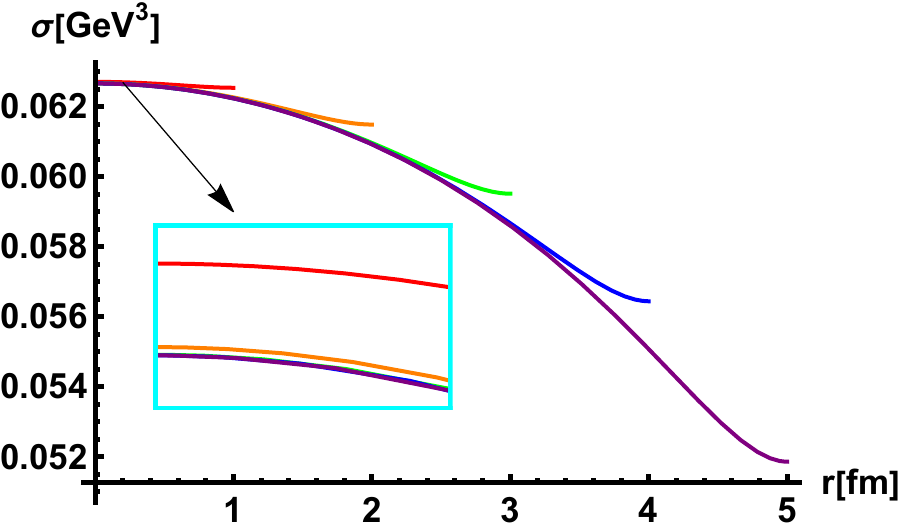}
\hspace*{1cm}
\includegraphics[width=0.45\textwidth]{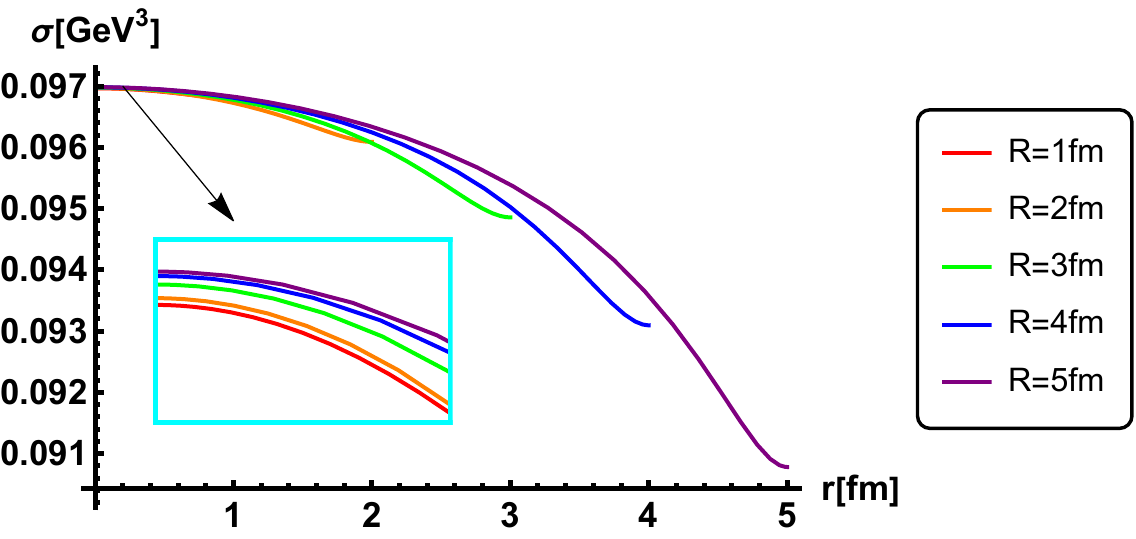}
\vskip -0.05cm \hskip -1.2 cm
\textbf{( a ) } \hskip 7.3 cm \textbf{( b )}
\caption{\label{fig:fs-N} The figures show the profiles of condensation for different radii with NBC, $(m_{q},\upsilon_{3},\upsilon_{4})=(0,-3,8)$ and $\Omega=0.01$. Among them, Fig.(a) and Fig.(b) are at $T=170$ MeV and $T=50$ MeV, respectively.}
\end{figure}

At an angular velocity of 0.01 GeV and a temperature $T=170$ MeV close to the
phase transition temperature $T_{c}$, the condensation as a function of the radial coordinate $r$
for various sizes $R$ is shown in Fig.\ref{fig:fs-N}(a). 
It can be seen from the figure that the variation of radius $R$ has little modification on the condensation profile under NBC.
As the size of the system decreases, the condensation is enhanced both at the center and the edges. 
Therefore, it can be concluded that the decrease of the system radius catalyses the chiral symmetry breaking, 
similar to the periodic boundary condition in the NJL model.
The enlarged portion of Fig.\ref{fig:fs-N}(a) shows the value of condensation at the center as a function of radius $R$. 
From the Fig.\ref{fig:fs-N}(a), it can be obtained that the catalysis effect of finite size is pretty weak under NBC.

When the temperature of the system is apparently below the critical
temperature $T_{c}$, the catalysis effect of finite size will be
converted to inverse catalysis in case of NBC. At the angular velocity
of 0.01 GeV and temperature of 50 MeV, the condensation as a function
of the radial coordinate $r$ is plotted in Fig.\ref{fig:fs-N}(b).
As can be seen in Fig.\ref{fig:fs-N}(b), the condensation diminishes with the decrease of the system size except near the edges. 
The behavior at the edges is from the role of boundary conditions. 
For the enlarged portion of Fig.\ref{fig:fs-N}(a) and (b),
it is clear that at the center, the condensation value of the smaller size is higher than that of the larger size at high temperature, whereas the opposite
is true at low temperature. The condensation
value at the center becomes smaller as the system size decreases,
which is the opposite of what happens at high temperature. 
Furthermore, no matter the catalysis effect or inverse catalysis effect of the finite size, 
they are quite weak under NBC and have little change on the condensation value.

From the results in Fig.\ref{fig:fs-N},
it is found that for the condensation at the center, the finite size
exhibits two totally opposite effects, i.e., catalysis at high temperature
and inverse catalysis at low temperature, in case of NBC. It can be
further deduced that the condensation value at the center may strengthen
or weaken as the system size changes due to the combination of the
two effects when the temperature is neither high nor low.

\subsection{Phase transition induced by finite size}

In this subsection, the effect of finite size on chiral condensation
is studied with DBC at the edges. Instead of the choices in the previous
subsection, the five radii $R$ are selected here as 2.2, 2.4, 3, 4
and 5 fm. The reason for such choices is that for
the radius $R=1$ fm, the system does not have inhomogeneous chiral condensation
at the temperature of 170 MeV and the angular velocity of 0.01 GeV with $(m_{q},\upsilon_{3},\upsilon_{4})=(0,-3,8)$.

\begin{figure}
\includegraphics[width=0.45\textwidth]{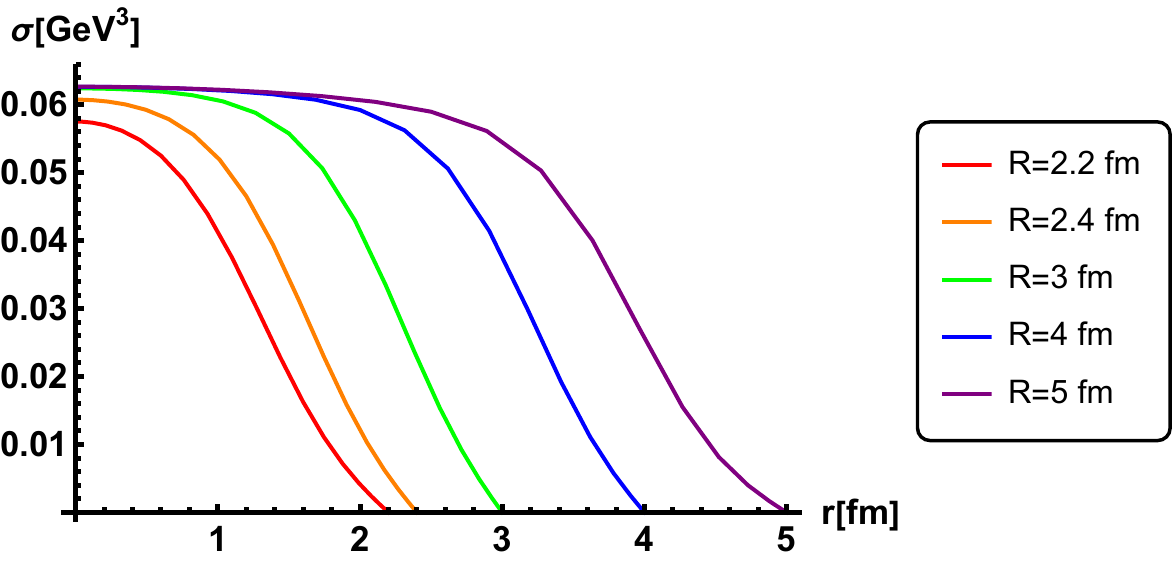}
\hspace*{1cm}
\includegraphics[width=0.34\textwidth]{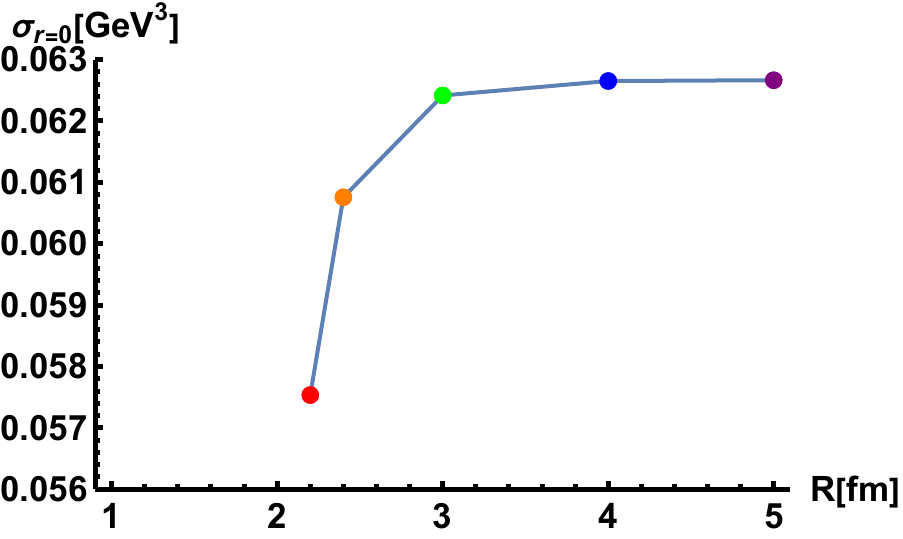}
\vskip -0.05cm \hskip 0 cm
\textbf{( a ) } \hskip 9.3 cm \textbf{( b )}
\caption{\label{fig:fs-D-ic} Fig.(a) shows the results of chiral condensation with different radii $R$, at $T=170$ MeV and $\Omega=0.01$ GeV with DBC and $(m_{q},\upsilon_{3},\upsilon_{4})=(0,-3,8)$. Fig.(b) shows the value of condensation at the center as a function of the radius $R$.}
\end{figure}

The chiral condensation as a function of the radial coordinate $r$ with
DBC is presented in Fig.\ref{fig:fs-D-ic}. Fig.\ref{fig:fs-D-ic}(a)
shows the condensation of the five radii selected and Fig.\ref{fig:fs-D-ic}(b)
gives the value of condensation at the center as a function of radius $R$. 
From the figure, it can be concluded
that the condensation in all spaces diminishes as the size of the
system decreases and the value of it near the center has a significant
decrease when the radius of the system is less than 3 fm.
This suggests that, for DBC, the effect of finite size on condensation
is inverse catalysis. Thus, it can be conjectured that when the system
size is small enough, the condensation will not have inhomogeneous
solutions, and the results of numerical calculations prove it. When
setting the radius $R=1$ fm, the condensation of the full
space is 0, i.e., the system is restored to the chiral symmetric phase.
So it can be further postulated that as the size decreases, the system
will suffer a phase transition from the inhomogeneous phase (its presence
at a point in space where the chiral condensation is not zero) to
the homogeneous phase (chiral symmetric phase). 

\begin{figure}
\includegraphics[width=0.38\textwidth]{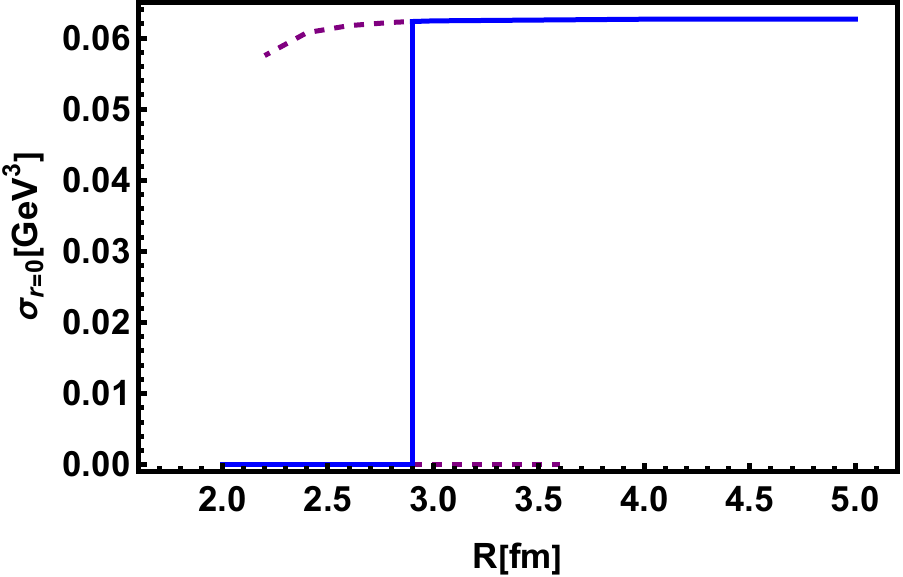}
\hspace*{1cm}
\includegraphics[width=0.38\textwidth]{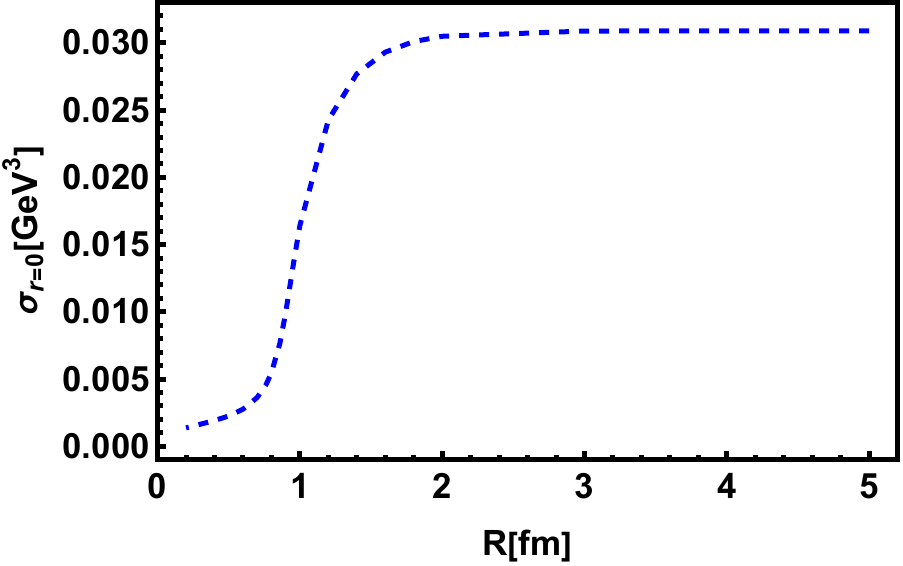}
\vskip -0.05cm \hskip 1 cm
\textbf{( a ) } \hskip 6.8 cm \textbf{( b )}
\vskip 0.5cm
\includegraphics[width=0.38\textwidth]{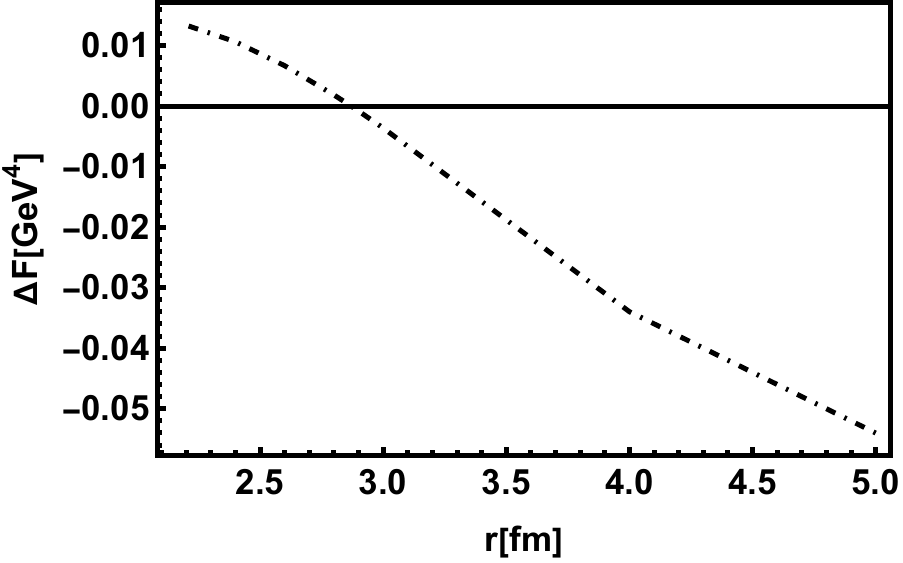}
\hspace*{1cm}
\vskip -0.05cm \hskip 0 cm \textbf{(c) }
\caption{\label{fig:fs-D-fd} Fig.(a) and (b) show the results of
chiral condensation at the center of the system as a function of radius
$R$. Here, Fig.(a) shows the results with $(m_q,\upsilon_{3},\upsilon_{4})=(0,-3,8)$ at $T=170$ MeV and $\Omega=0.01$ GeV, while Fig.(b) is the results with $(m_q,\upsilon_{3},\upsilon_{4})=(7{\rm MeV},0,8)$ at $T=115$ MeV and $\Omega=0.01$. 
Fig.(c) shows the difference in free energy between the inhomogeneous
and homogeneous phases as a function of radius $R$ for $(m_q,\upsilon_{3},\upsilon_{4})=(0,-3,8)$.}
\end{figure}

This postulation that the reduction of the system size induces a phase
transition can be verified by comparing the difference between the free
energy of the inhomogeneous and homogeneous phases. The free energy
difference between the two phases is shown in Fig.\ref{fig:fs-D-fd}(c), and it can be noted that the difference is
positive with $R_c\lesssim 2.8$ fm and negative with $R_c\gtrsim 2.8$ fm.
Therefore, the homogeneous phase is more stable when the system size
is less than the critical size $R_c\simeq 2.8$ fm, and conversely, the inhomogeneous
phase is preferred. Since the condensation is maximum at the center,
the value here is chosen to determine whether the system is in the
symmetric phase. The value of the condensation at the center is shown
as a function of the size $R$ in Fig.\ref{fig:fs-D-fd}(a),
where the blue solid and purple dashed lines indicate the stable and
metastable phases, respectively. When the parameters of the potential
term Eq.(\ref{eq:potential-1}) are chosen as $(m_{q},\upsilon_{3},\upsilon_{4},)=(7{\rm MeV},0,8)$,
i.e., the case of crossover, the phase transition induced by the finite
size turns into crossover and the pseudo-critical size is around $R_c\simeq 1$ fm as in Fig.\ref{fig:fs-D-fd}(b). Consequently, the order of the phase transition
induced by the finite size is determined by the form of the potential
term Eq.(\ref{eq:potential-1}). For the case where the potential parameters are the first-order phase transition, the finite size induces a first-order
phase transition, and so forth. In summary, the system size $R$ acts
similarly to the inverse of temperature $1/T$, inducing a phase transition
from inhomogeneous to homogeneous phase.

\section{phase diagram with the angular velocity $\Omega$\label{sec:phase-diagram}}

In this section, the phase diagram of chiral transition on the temperature-angular
velocity plane will be discussed. It should be emphasized that the
condensation values are different in all points of the space due to
the rotation of the system. In order to study the $T-\Omega$ phase
diagram, following the previous section, the condensation value at
the center $\sigma_{r=0}$ is chosen as the order parameter of the
system. In addition, it is known from the above that the finite
size has an impact on the phase diagram, especially when the DBC is
selected. Hence, the size of the system is fixed to 2 fm or 4 fm,
which are the typical sizes of the fire balls created in the current heavy ion collisions, in order to compare the finite
size effect. 

\begin{figure}
\includegraphics[width=0.38\textwidth]{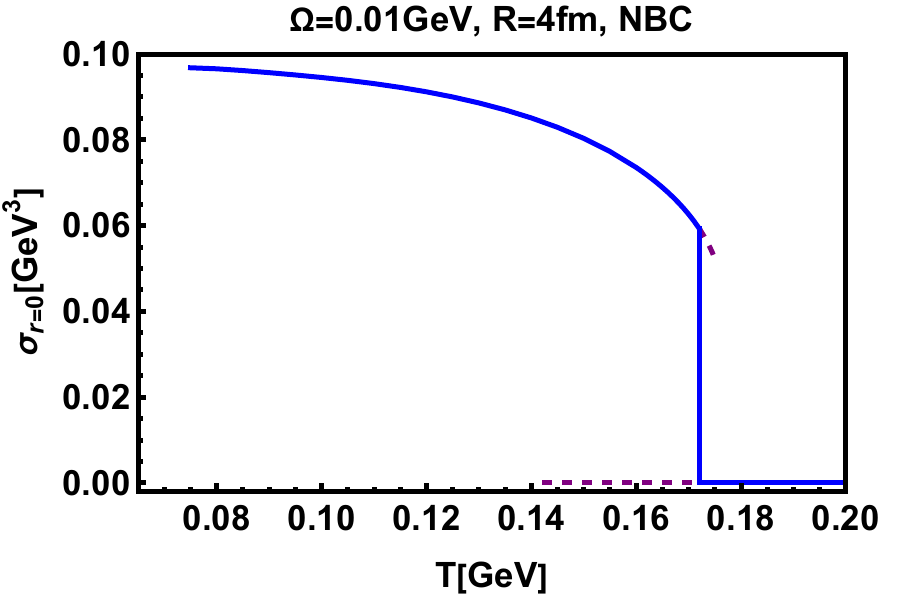}
\hspace*{1cm}
\includegraphics[width=0.38\textwidth]{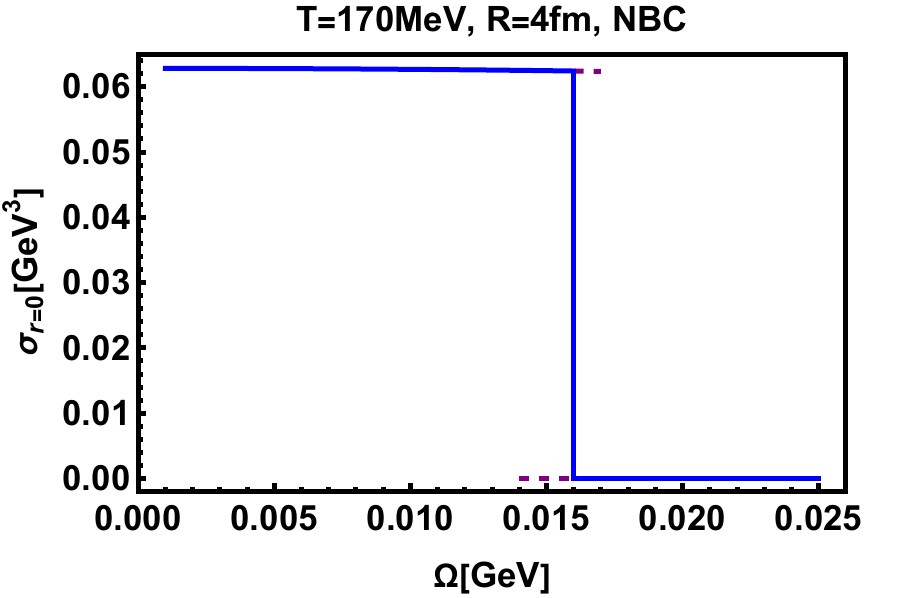}
\vskip -0.05cm \hskip 1 cm
\textbf{( a ) } \hskip 6.8 cm \textbf{( b )}
\vskip 0.5cm
\includegraphics[width=0.38\textwidth]{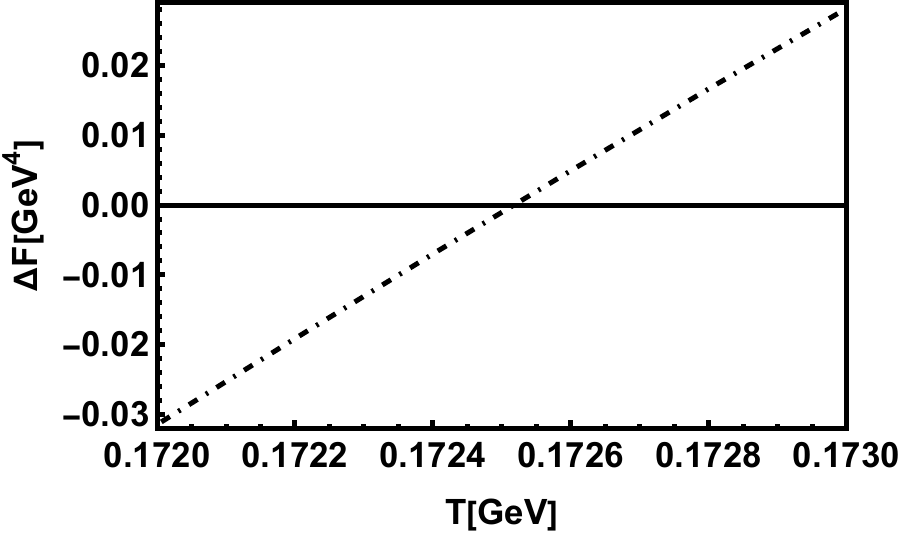}
\hspace*{1cm}
\includegraphics[width=0.38\textwidth]{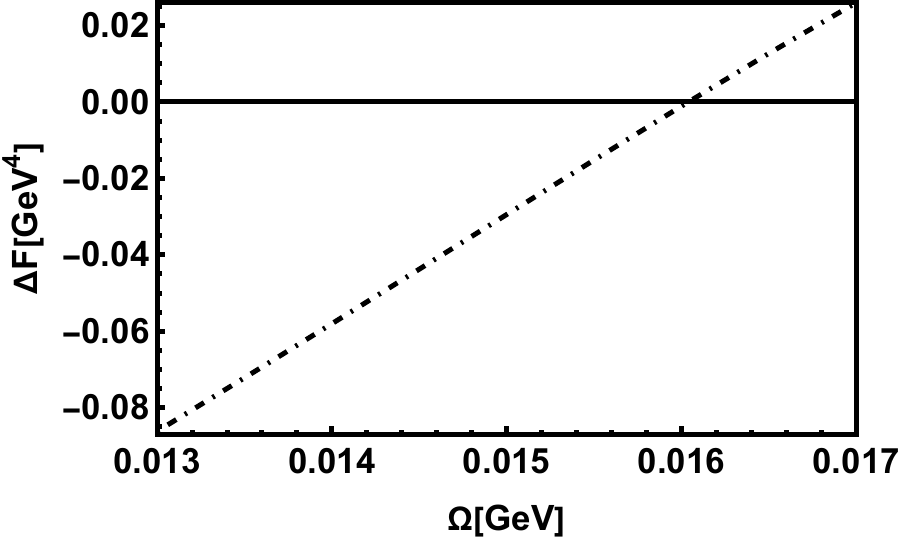}
\vskip -0.05cm \hskip 1 cm
\textbf{( c ) } \hskip 6.8 cm \textbf{( d )}
\caption{\label{fig:sig-T-O} In Fig.(a) and (b), the chiral condensation at the center as a function of temperature $T$ or angular velocity $\Omega$ are shown. Here, the solid blue line indicates
the stable state while the purple dashed line denotes the metastable
state. Fig.(c) and (d) shows the difference in free
energy between the inhomogeneous and homogeneous phases as a function
of temperature $T$ or angular velocity $\Omega$.}
\end{figure}

In Sec.\ref{sec:chi-rot}, it has been shown that with the parameters $(m_q,\upsilon_{3},\upsilon_{4})=(0,-3,8)$, when the temperature or angular velocity increases to certain critical values, the inhomogeneous solutions becomes disfavored and the chiral symmetry is restored everywhere. Thus, the system undergoes the chiral phase transition at those critical values. Further, the critical
temperature $T_{c}$ and the critical angular velocity $\Omega_{c}$
of the phase transition can be determined by the difference of the
free energy between the inhomogeneous and homogeneous phases. The
free energy difference and the condensation at the center
as a function of temperature $T$ or angular velocity $\Omega$ are
shown in Fig.\ref{fig:sig-T-O}. In particular, Fig.\ref{fig:sig-T-O}(a) shows the condensation at the center as a function of temperature $T$ for $\Omega= 0.01$ GeV and $R= 4$ fm with NBC; Fig.\ref{fig:sig-T-O}(b) shows condensation at the center as a function of angular velocity for $T= 170$ MeV and $R= 4$ fm with NBC. And the free energy
difference as a function of temperature and angular velocity is shown
in Fig.\ref{fig:sig-T-O}(c) and (d), respectively.
Here, the blue solid and purple dashed lines denote the stable and metastable
phases, respectively. As can be seen from the figure, the critical
temperature $T_{c}\simeq 172.5$ MeV for $\Omega=0.01$ GeV drops about 1 MeV compared to that
without rotation, which is not a big
modification. As for the fixed temperature, the condensation at the
center varies little for small angular velocities and the system returns
to homogeneous phase when the critical angular velocity $\Omega_{c}\simeq0.016$GeV
is reached.

\begin{figure}
\includegraphics[width=0.38\textwidth]{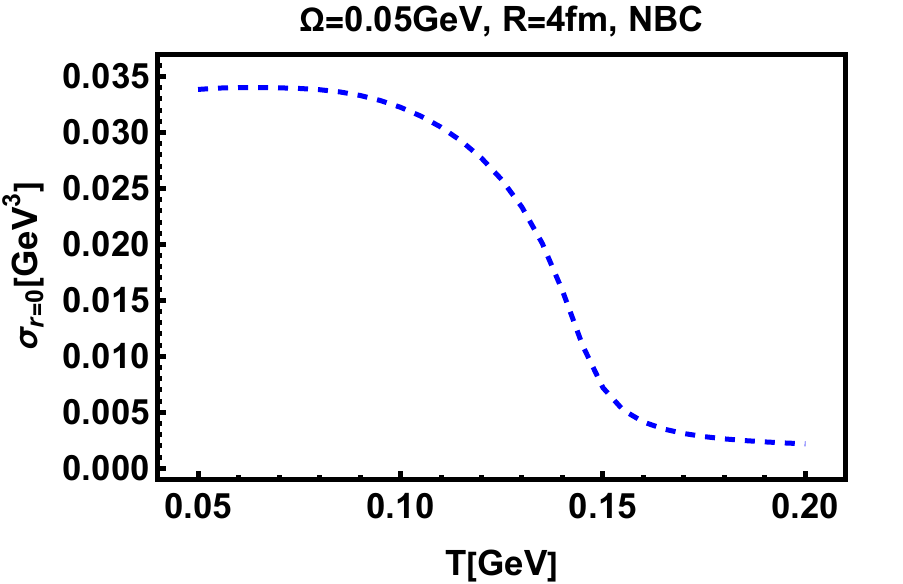}
\hspace*{1cm}
\includegraphics[width=0.38\textwidth]{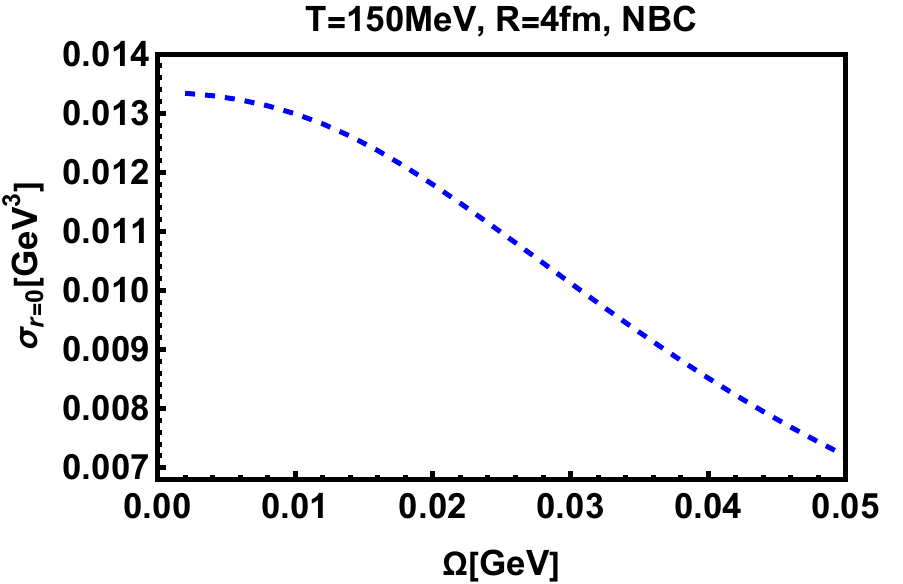}
\vskip -0.05cm \hskip 1 cm
\textbf{( a ) } \hskip 7. cm \textbf{( b )}
\caption{\label{fig:sig-T-O-2} Fig.(a) and (b) represent the chiral
condensation values at the center as a function of temperature $T$
and angular velocity $\Omega$, respectively, with NBC and $(m_q,\upsilon_{3},\upsilon_{4})=(7{\rm MeV},0,8)$.}
\end{figure}

When the parameters of the model are set to $(m_q,\upsilon_{3},\upsilon_{4})=(7{\rm MeV},0,8)$, the phase transition
of the chiral condensation at the center is also converted to crossover
and is shown in Fig.\ref{fig:sig-T-O-2}. In Fig.\ref{fig:sig-T-O-2}(a) shows the condensation at the center as a function of temperature
at $\Omega=0.05$GeV and $R=4$ fm
with NBC, and Fig.\ref{fig:sig-T-O-2}(b) shows the condensation at the center as a function of angular velocity at $T=150$ MeV and
$R=4$ fm with NBC. It is easy to see that $\sigma_{r=0}$ decreases slowly at low $T$ and small $\Omega$, and it decreases fast near the transition temperature, showing a typical behavior of crossover in a narrow range of $T$ or $\Omega$. Usually, one can define the pseudo-transition temperature where the slope reaches its maximum, i.e. $\frac{d^2 \sigma_{r=0}}{dT^2}$ or  $\frac{d^2 \sigma_{r=0}}{d\Omega^2}=0$.

\begin{figure}
\includegraphics[width=0.55\textwidth]{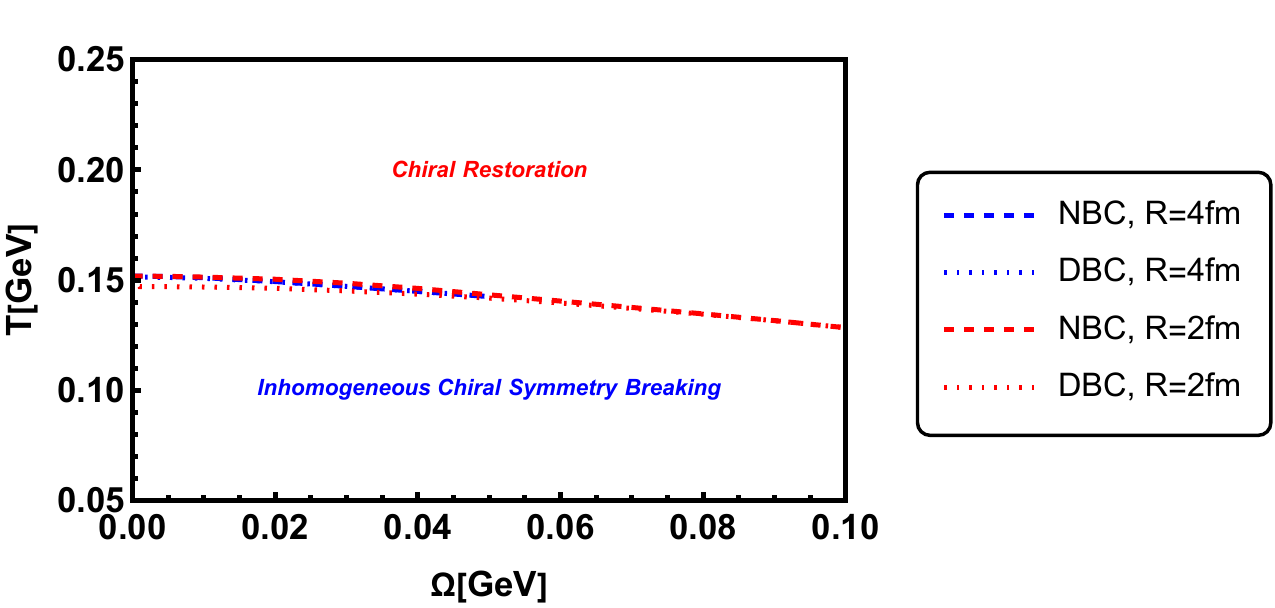}

\caption{\label{fig:TOpd} The $T-\Omega$ phase diagram at fixed radius $R$
is shown in the figure. Here, the red and blue lines correspond to radii of $R=2$ fm and 4 fm, respectively, and the dashed and dotted lines represent NBC and DBC, respectively.}
\end{figure}

With the above discussion, for each temperature one can get a critical $\Omega$ and the phase diagram of the chiral phase transition in $T-\Omega$ space
can be obtained. 
In heavy ion collisions, the rotation of hot dense matter may have measurable consequences for chiral symmetry restoration/breaking. Therefore, the study of the $T-\Omega$ phase diagram of QCD matter can demonstrate the possible influences.
In the case of $N_{f}=2$, the
chiral phase transition from the lattice QCD simulation is crossover,
so the model parameters are chosen to be $(m_{q},\upsilon_{3},\upsilon_{4})=(7{\rm MeV},0,8)$
in which the phase transition temperature is close to 150MeV and the
vacuum value of $\sigma$ is about ($0.320{\rm GeV})^{3}$. Furthermore,
as previously discussed, the finite size affects the phase diagram
and phase structure, so the radii of the system are chosen to be 4 fm
and 2 fm for comparison. In both cases, the value of
the uniform angular velocity is limited to 0.05GeV and 0.1GeV, in order not to violate the causality. The final results for the phase diagram are shown in Fig.\ref{fig:TOpd},
where the red line and the blue line correspond to the
radii $R=2$ fm and $R=4$ fm, respectively.
In the holographic model, the boundary conditions of the scalar field $\chi$ can be selected as
NBC and DBC at the edges, and the results are listed in Fig.\ref{fig:TOpd} with dashed line and dotted line, respectively. As can be seen from the figure, the critical temperarure would decrease with the angular velocity, and such a  $T-\Omega$ phase diagram is similar to the $T-\mu$ phase diagram,
which is also consistent with the previous conclusion that the angular
velocity is analogous to the chemical potential. If the system size
decreases, the shifting of the phase boundary depends on the choice
of boundary conditions, for NBC the boundary line moves outward, while
for DBC the boundary line moves inward. This is consistent with the
effect of finite size on condensation for different boundary conditions,
i.e., near the critical temperature, NBC and DBC behave as catalysis
and inverse catalysis, respectively. For NBC, the phase boundary line
is slightly higher than the case of DBC. In summary, the $T-\Omega$
phase diagrams are only slightly different even when considering the
boundary conditions and system size. If we consider the angular velocity
of about 0.01 GeV produced by heavy ion collision, the temperature
of phase transition is about 128 MeV, which shows a significant effect of the rotation.

In Ref.\citep{Jiang2016}, the first-order phase transition occurs
at low temperature high angular velocity with critical end point located
at $T_{CEP}\simeq0.02$GeV and $\Omega_{CEP}\simeq0.644$GeV. 
Since the phase
structure depends on the radius $R$, to produce an angular velocity
of 0.65 GeV either limit the system size to less than 1.5${\rm GeV}^{-1}$
or consider the angular velocity of the vortex configuration. For
the former, the radius of the system is so small that the finite size
has a large impact on the final result. For the latter, the uncertainty
of the phase diagram for different configurations cannot be estimated.
For these reasons, phase diagram for large angular velocity is not
included in this paper.

\section{conclusion and discussion}
\label{sec:sum}

In this paper, the configuration of chiral condensation under rotation considering finite size and boundary conditions,
and the temperature-angular velocity phase diagram are investigated and discussed in the holographic QCD model. In the holographic model, the rotation always exhibits a suppression effect determined by $\Omega r$, which is consistent with Ref.\citep{Wang2019}, while the centrifugal-like effect of Ref.\citep{Wang2019} does not appear.
In addition, the influence of Neumann and Dirichlet boundary conditions on the profile of condensation is mainly focused on the edges. 
For the finite-size effect, it depends on the choice of boundary conditions. For NBC, the condensation at the center is catalysis at high temperature and
inverse catalysis at low temperature, as the radius of the system
decreases. For DBC, chiral condensation is inverse catalysis and the
phase transition is induced by decreasing the size. $T-\Omega$
phase diagrams are obtained for different boundary conditions and
system sizes, and the differences between them are not significant.
At the angular velocity of 0.01 GeV, which is typical in heavy-ion
collisions, the critical temperature of the phase transition is about
128 MeV.

In the holographic model, the condensation under rotation exhibits the local suppression effect, regardless of the boundary conditions or rotational velocity distribution that are selected. Three cases of rotational velocity distribution are discussed: (i) constant angular velocity, (ii) angular velocity of the vortex structure and (iii) angular velocity concentrated at the center. For case (i), the shape of the condensation as a function of the radial coordinate $r$ is similar to a swelling or a drum, for case (ii) the condensation has a minimum value at the center of the vortex, and for case (iii) the condensation decreases slightly near the center. The condensation does not display the centrifugal-like effect as shown in Ref.\citep{Wang2019}, although case (iii) or small angular velocities are given as input. The possible reason is that the rotation is introduced through the polar part of the gauge field, which acts similarly to the chemical potential. The condensation with DBC has a plateau profile near the center, which is consistent with Ref.\citep{Ebihara2017} and confirms the validity of the local density approximation. However, near the edge $R$, the model does not show the oscillating behavior, which is inconsistent with \citep{Ebihara2017}.

Different from the configuration of inhomogeneous condensation, the finite size effect with uniform angular velocity depends on the choice of boundary conditions, similar to that discussed in Ref.\citep{Xu2020}. For NBC, at slightly below the critical temperature, the decreasing size enhances the condensation value in the full space, especially at the center, which behaves as the catalysis effect. However, away from the critical temperature, the finite size effect transforms into inverse catalysis, which is analogous to the behavior of magnetic fields. For DBC, the finite size shows the inverse catalysis effect at any temperature. And, as the size decreases, the system exhibits phase transitions from the chiral broken phase to the chiral restored phase. The order of the phase transition depends on the form of the potential term in the 5D action, if the parameters of the potential term are chosen as first-order phase transition then the phase transition is first-order, and so forth.

The temperature-angular velocity phase diagram of chiral condensation has been studied in the holographic model. Since the condensation is inhomogeneous, the condensation at the center is taken as the order parameter of the system. In consideration of the fact that both finite size and boundary conditions can influence the profile of condensation, we selected four cases with radii of 2 fm or 4 fm and NBC or DBC, as a comparison. The results show that although the phase lines of the four cases do not overlap, the differences between them are negligible. The temperature-angular velocity phase diagram at large angular velocity are not discussed because the size of the QGP is taken into account. If the vortex structure similar to the Abrikosov lattice is achieved in the holographic QCD model, the exploration of the phase structure at higher rotational speed will be possible.

\begin{acknowledgments}
This work is supported in part by the China Postdoctoral Science Foundation under Grant No. 2021M703169, the National Natural Science Foundation of China (NSFC) Grant  Nos 11725523, 11735007, 11805084 and  the Strategic Priority Research Program of Chinese Academy of Sciences under Grant Nos XDB34030000 and XDPB15, the start-up funding from University of Chinese Academy of Sciences(UCAS), the Fundamental Research Funds for the Central Universities,and Guangdong Pearl River Talents Plan under Grant No. 2017GC010480.
\end{acknowledgments}

\bibliographystyle{unsrt}
\bibliography{2-ref}

\end{document}